\pdfoutput=1
\documentclass[11pt,a4paper]{article}

\usepackage[utf8]{inputenc}
\usepackage[T1]{fontenc}
\usepackage{lmodern}
\usepackage[margin=1in]{geometry}
\usepackage{microtype}
\usepackage{amsmath,amssymb,amsfonts,amsthm}
\usepackage{algorithm}
\usepackage{algorithmic}
\usepackage{graphicx}
\usepackage{booktabs}
\usepackage{tabularx}
\usepackage{multirow}
\usepackage{array}
\usepackage{xcolor}
\usepackage{enumitem}
\usepackage{cite}
\usepackage{url}
\usepackage{fancyhdr}
\usepackage{listings}
\usepackage{caption}
\usepackage[section]{placeins}
\usepackage[colorlinks=true,linkcolor=blue!55!black,citecolor=blue!55!black,urlcolor=blue!55!black]{hyperref}

\hypersetup{
  pdftitle={BTAE: Bottom-Up Traceable Agentic Engineering via Deterministic Knowledge Graphs and Four-Dimensional Epistemic Governance},
  pdfauthor={Zhengshuai Yang, Changyun Li},
  pdfkeywords={Agentic software engineering, engineering knowledge graph, bidirectional traceability, architectural fitness functions, evidence-based governance, coding agents}
}

\newtheorem{definition}{Definition}
\newtheorem{proposition}{Proposition}
\newtheorem{lemma}{Lemma}

\fancypagestyle{plain}{
  \fancyhf{}
  \fancyfoot[C]{\footnotesize\thepage}
  
}

\title{\textbf{BTAE: Bottom-Up Traceable Agentic Engineering via Deterministic Knowledge Graphs and Four-Dimensional Epistemic Governance}}

\author{%
  Zhengshuai Yang \qquad Changyun Li\thanks{Corresponding author. Preprint. This work was supported in part by the Hunan Provincial Natural Science Foundation under Grant 2026JJ80034, in part by the New Engineering Research and Practice Project of the Hunan Provincial Department of Education under Grant 202503000046, and in part by the China Scholarship Council under Grant CXXM2510173682.}\\
  \small College of Artificial Intelligence, Hunan Agricultural University, Changsha 410128, China\\
  \small \texttt{yhvguk@stu.hunau.edu.cn}, \texttt{362330193@qq.com}
}

\date{}

\begin{document}
\sloppy\emergencystretch=3em
\maketitle

\begin{abstract}
Autonomous coding agents accelerate localized patch synthesis, yet sustaining repository-level software evolution across multi-turn edits requires preserving verifiable links among specifications, architectural boundaries, source symbols, and test executions. Existing agent scaffolds and specification-driven toolkits struggle in long-horizon maintenance because they conflate self-reported metadata, static file checks, and historical test outcomes with currently valid evidence, allowing stale test runs, silent architectural drift, risk self-downgrades, and unverified host claims to pass completion gates while flooding agent contexts with irrelevant repository files. Our key insight is that reliable agentic engineering must decouple structural facts from governance authority by tracking an orthogonal four-dimensional epistemic state---design approval, implementation mapping, verification execution, and Assumption-Based Truth Maintenance (ATMS)-sliced defeasible validity---for every logical engineering entity. Guided by this principle, we present \textbf{Bottom-Up Traceable Agentic Engineering (BTAE)} and its reference \textbf{Engineering Protocol Engine (EPE)}, which pairs an atomic, Directory-Merkle content-addressed \textbf{Engineering Knowledge Graph (EKG)} with a fail-closed \textbf{Engineering Governance Plane (EGP)}. Without requiring external graph or vector servers, EPE deterministically reconstructs entity-to-symbol mappings and module import boundaries, assembles stratified \texttt{C0}--\texttt{C3} task contexts with fail-closed caching, derives policy-enforced risk floors and global protected paths from EKG blast radii, enforces Ed25519-signed design approvals and time-bounded Architecture Decision Record (ADR) waivers, and binds subprocess JUnit verification artifacts to SHA-256 input manifests, ATMS assumption slices separating \texttt{undercut} from \texttt{rebutted} failures, and hash-chain audit logs. Evaluated on the self-hosted 7,338-LOC EPE repository (426 EKG nodes, 550 edges, 368 automated tests) and a 620-trial randomized fault-injection benchmark (14 adversarial families totaling 560 trials plus 60 valid workflows), EPE achieves 10/10 byte-identical cold graph reconstructions in 132.2\,ms median latency (57.5\,ms median incremental under racy-safe stat caching, $2.30\times$ median speedup), reduces conditional false acceptance from 49.5\% in a strong Git+CI+static stack (\textbf{B3}) and 83.8\%--100.0\% in narrower baselines to \textbf{11.1\%} (95\% CI 8.7--13.9) under anchored policy boundaries (25.4\% locally with risk floors, 32.5\% unanchored) with 0/60 false rejections on standard compliant workflows, compresses task context payloads by \textbf{97.6\%} relative to full-repository loading (1.98\,ms cached latency on a non-truncated bundle), and shrinks evidence invalidation slices by \textbf{78.9\%} on average, while quantifying its 15/15 fail-closed cost on standard-library lazy loading and explicitly delineating its local policy-boundary limits.
\end{abstract}


\vspace{0.6em}
\hrule
\vspace{0.6em}

\section{Introduction}
\label{sec:intro}

Autonomous coding agents have progressed rapidly from single-function synthesis to repository-level issue resolution across real-world codebases~\cite{jimenez2024swebench,yang2024sweagent,wang2025openhands,zhang2024autocoderover}. In continuous software maintenance, however, shipping an isolated patch is only half of the engineering problem: each modification must remain traceable to an explicit requirement or contract, respect module dependency boundaries, stay within an authorized write scope, and leave behind verification evidence that remains valid for the exact state of the working tree. When multiple agent turns accumulate without structural governance, short-term coding velocity compounds into long-term software aging and architectural erosion~\cite{parnas1994aging,perry1992architecture}: agents edit against outdated contracts, introduce hidden cross-layer imports, and declare tasks ``complete'' on the basis of stale or unverified test claims.

Classical software engineering addresses design-to-code divergence through requirements traceability~\cite{gotel1994traceability,clelandhuang2014traceability}, software reflexion models~\cite{murphy1995reflexion,murphy2001reflexion}, and automated trace-link recovery between architecture documentation and source code~\cite{keim2024transarc,fuchss2025lissa}. More recently, repository knowledge graphs such as RepoDoc~\cite{xu2026repodoc} extract AST-level call and import structures to reduce context token inflation during documentation generation. Yet these approaches are predominantly descriptive rather than prescriptive: they reconstruct or summarize existing codebases post hoc, treat probabilistic or heuristic links symmetrically with human-authored specifications, and lack fail-closed governance gates to constrain what an autonomous coding agent may modify or finalize during live execution.

Conversely, emerging specification-driven agent environments~\cite{github2025speckit,aws2025kiro} and contract-grounded test generation pipelines~\cite{meyer1992dbc,tufano2025contracts} reintroduce upfront specifications before code edits, but leave two structural vulnerabilities unresolved in iterative maintenance. First, mechanical one-file-per-document conventions break under routine refactoring and force rigid waterfall workflows onto heterogeneous tasks such as urgent bugfixes, baseline-preserving refactorings, and exploratory prototypes. Second, and more fundamentally, existing agent loops suffer from \emph{epistemic conflation}: they collapse an entity's lifecycle into a single boolean or frontmatter flag, conflating an agent's self-reported declaration (\texttt{design\_status:\ approved}), a static symbol check, or a historical test pass (\texttt{result:\ passed}) with current evidentiary applicability (\texttt{validity:\ valid}) after subsequent source, policy, or environment mutations.

\begin{figure}[t]
  \centering
  \includegraphics[width=\textwidth]{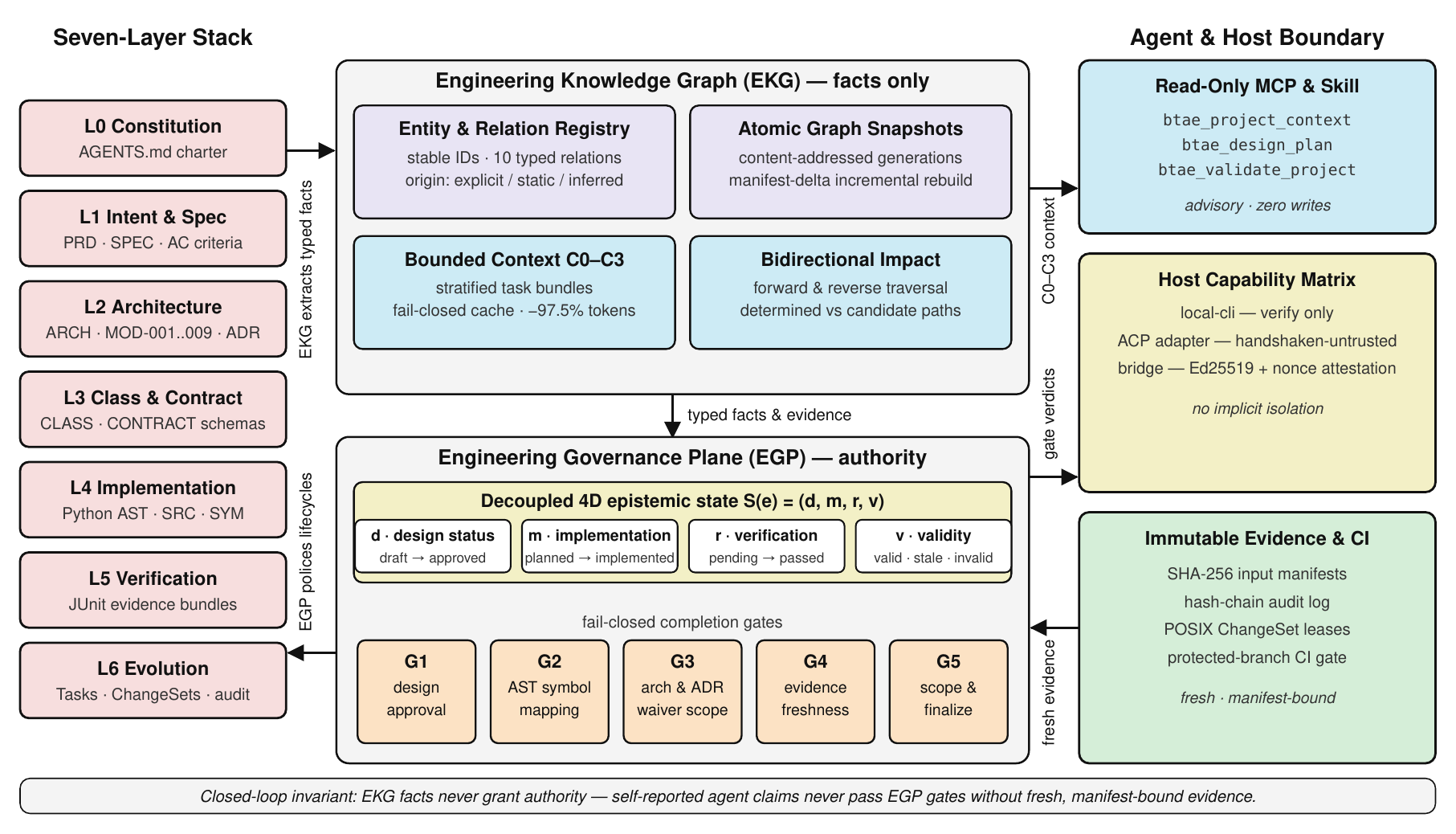}
  \caption{Overview of Bottom-Up Traceable Agentic Engineering (BTAE) and the Engineering Protocol Engine (EPE). Left: the non-waterfall seven-layer responsibility stack (\texttt{L0}--\texttt{L6}). Center: strict separation between the fact-only Engineering Knowledge Graph (EKG) and the authority-enforcing Engineering Governance Plane (EGP) operating over the decoupled 4D epistemic state $\mathcal{S}(e)$. Right: explicit trust boundaries across read-only MCP/Skill planning, capability-attested host adapters, and manifest-bound CI evidence.}
  \label{fig:architecture}
\end{figure}

To eliminate this epistemic conflation, we introduce \textbf{Bottom-Up Traceable Agentic Engineering (BTAE)} and its local deterministic reference engine, the \textbf{Engineering Protocol Engine (EPE)}. As illustrated in Figure~\ref{fig:architecture}, BTAE organizes software responsibility across a non-waterfall seven-layer stack (\texttt{L0} Constitution through \texttt{L6} Evolution) while strictly separating two horizontal facilities: an \textbf{Engineering Knowledge Graph (EKG)} that records provenance-typed facts, and an \textbf{Engineering Governance Plane (EGP)} that adjudicates lifecycle transitions and completion gates. Rather than tracking a single status flag, EPE evaluates every logical entity under an orthogonal four-dimensional epistemic state $\mathcal{S}(e) = (d, m, r, v)$ spanning design approval ($d$), AST-verified implementation mapping ($m$), JUnit-parsed verification outcome ($r$), and ATMS-sliced defeasible validity ($v$, with coarse disk status in $\{\texttt{valid}, \texttt{stale}, \texttt{invalid}, \texttt{unknown}\}$ refined into Pollock-style \texttt{undercut} vs.\ \texttt{rebutted} defeaters). Consequently, an entity whose design is \texttt{approved}, implementation is \texttt{implemented}, and historical test run \texttt{passed} is automatically downgraded to \texttt{stale}/\texttt{undercut}---and blocked at finalization---the instant a single byte changes in its ATMS assumption slice or verification policy.

Building on this separation of facts and authority, EPE supports practical day-to-day agentic maintenance without requiring external graph databases, vector stores, or cloud LLM judges---offering a strong on-premise privacy advantage because proprietary source code never leaves the local host. During task preparation, a read-only Model Context Protocol (MCP)~\cite{anthropic2024mcp} server (\texttt{btae-mcp}) and companion agent skill assemble stratified \texttt{C0}--\texttt{C3} context bundles prioritized by mandatory charter rules, bidirectional EKG neighborhoods, and explicit implementation/test mappings, while refusing to cache truncated or incomplete bundles. During modification and review, EPE performs direction-aware impact propagation that distinguishes \texttt{determined} static/explicit paths from \texttt{candidate} inferred paths, enforces policy-derived minimum risk floors and global protected paths from EKG blast radii so agents cannot silently downgrade high-impact tasks, and explicitly flags dynamic-import blind spots (\texttt{EPE-AST-DYNAMIC}). Across execution boundaries, EPE models host capabilities conservatively: local CLI invocations and unattested Agent Client Protocol (ACP) handshakes remain \texttt{handshaken-untrusted}, whereas privilege elevation across host bridges requires Ed25519 SSH signatures bound to per-request nonces and configured trust roots.

We evaluate BTAE and EPE through self-hosted governance of the 7,338-LOC EPE codebase (17 logical design/contract entities, 54 managed Python source files across 9 logical modules plus the \texttt{mcp/} entrypoint, 426 EKG nodes, 550 edges---4 explicit \texttt{realizes} and 546 static \texttt{depends\_on}---and 368 unit, integration, and acceptance tests) alongside a 620-trial randomized adversarial fault-injection ablation (14 families totaling 560 adversarial trials plus 60 valid workflows) and synthetic scaling benchmarks up to 400 Python files. Across all evaluations, EPE reconstructs byte-identical graph digests across 10/10 cold runs in 132.2\,ms median latency ($2.30\times$ median speedup under single-file incremental updates with 50\,ms racy-safe stat caching and Directory-Merkle digests), reduces conditional false acceptance from 49.46\% in a strong Git+CI+static stack (\textbf{B3}) and 83.75\%--100.00\% in narrower baselines to \textbf{11.07\%} (95\% CI 8.73--13.94) under anchored policy boundaries (25.36\% with local risk floors, 32.50\% unanchored) with 0/60 false rejections on standard compliant workflows (contrasted with 15/15 conservative rejections on probed standard-library lazy loading under \texttt{unknown\_is\_blocking = true})---every silent acceptance confined to two documented boundary classes---reduces task context payloads by \textbf{97.57\%} on average relative to full-repository loading (8/8 cache invalidation probes), and shrinks ATMS evidence slices by \textbf{78.92\%} while preserving 10/10 disjoint-leaf validity. Specifically, this paper makes four contributions:
\begin{itemize}[leftmargin=1.6em,itemsep=2pt,topsep=4pt]
  \item \textbf{BTAE Methodology and Four-Dimensional Epistemic Model:} We formulate a non-waterfall, entity-centric agentic engineering methodology that decouples design status, implementation mapping, verification execution, and ATMS-sliced defeasible validity across four workflow entrypoints (\texttt{feature}, \texttt{bugfix}, \texttt{refactor}, and \texttt{exploration}).
  \item \textbf{Deterministic Atomic EKG and Stratified Context/Impact Engine:} We design file-authoritative, lock-protected generation snapshots (\texttt{.engineering/generations/}) with Directory-Merkle inclusion proofs, 50\,ms racy-safe stat-delta incremental rebuilds, bounded generation GC, uncertainty-aware bidirectional impact analysis, and bounded \texttt{C0}--\texttt{C3} context resolution.
  \item \textbf{Fail-Closed Governance Gates, Risk Floors, and Explicit Host Trust:} We implement cryptographic Ed25519 design-revision approvals, policy-derived blast-radius risk floors and global protected paths, ADR-scoped and deadline-enforced fitness exceptions, JUnit-verified manifest-bound evidence with \texttt{undercut}/\texttt{rebutted} defeater diagnosis, append-only SHA-256 hash-chain audit logs, POSIX ChangeSet leases, and nonce-attested host capability handshakes.
  \item \textbf{Self-Hosted Reference Engine and Reproducible Empirical Suite:} We release the open-source Python EPE implementation, 15 Draft 2020-12 JSON Schemas, read-only MCP/Skill integrations, and reproducible 6-regime conditional-FAR fault-injection and scaling benchmarks, while explicitly delineating the boundaries between local risk-floor checks, CI/OS policy anchoring, and future longitudinal human-team studies.
\end{itemize}

\section{Methodology: BTAE and the Engineering Protocol Engine}
\label{sec:method}

Bottom-Up Traceable Agentic Engineering (BTAE) governs multi-turn coding agents inside a version-controlled repository through a file-authoritative protocol implemented by the local Engineering Protocol Engine (EPE). As depicted in Figure~\ref{fig:architecture}, BTAE partitions engineering artifacts into a non-waterfall seven-layer responsibility stack (\texttt{L0} Constitution through \texttt{L6} Evolution) while enforcing a strict architectural boundary between two horizontal facilities: the \emph{Engineering Knowledge Graph (EKG)}, which extracts and indexes provenance-typed facts, and the \emph{Engineering Governance Plane (EGP)}, which evaluates authorization, lifecycle transitions, and completion gates. Below, Section~\ref{sec:method_model} formalizes the entity-centric knowledge representation and decoupled four-dimensional epistemic state; Section~\ref{sec:method_graph} details atomic content-addressed graph publication and manifest-delta incremental reconstruction; Section~\ref{sec:method_context} presents stratified \texttt{C0}--\texttt{C3} context assembly and uncertainty-aware bidirectional impact analysis; and Section~\ref{sec:method_governance} specifies cryptographic approvals, ADR-scoped fitness waivers, manifest-bound verification evidence, and host capability attestation.

\subsection{Entity-Centric Knowledge Model and Four-Dimensional Epistemic State}
\label{sec:method_model}

Mechanical documentation schemes that pair one Markdown file with every source file degrade rapidly during agentic maintenance because routine file splits, renames, and symbol moves invalidate document identity and inflate synchronization overhead. Conversely, tracking a single status flag per document conflates whether a design has been reviewed, whether its target symbols exist, whether its tests passed historically, and whether those test results still apply to the current working tree.

We represent repository knowledge around stable, non-reusable logical entity identifiers governed by 15 Draft 2020-12 JSON Schemas (strictly rejecting unknown properties and duplicate YAML keys). Each entity $e \in \mathcal{V}$ carries a typed prefix (\texttt{INTENT-}, \texttt{SPEC-}, \texttt{ARCH-}, \texttt{MOD-}, \texttt{COMP-}, \texttt{CLASS-}, \texttt{CONTRACT-}, \texttt{ADR-}, \texttt{TASK-}, \texttt{CHANGE-}, \texttt{TEST-}, or \texttt{EVIDENCE-}) supplemented by derived source (\texttt{SRC-}), symbol (\texttt{SYM-}), and execution-run (\texttt{RUN-}) locators. Rather than relying on filename conventions, design entities bind to code and tests through explicit many-to-many \texttt{implementation[]} (\texttt{path} and optional qualified \texttt{symbol}) and \texttt{verification[]} (\texttt{test\_id} and \texttt{path}) mappings. Directed edges across $\mathcal{V}$ follow ten typed relations:
\begin{align*}
  \mathcal{R} = \{&\texttt{contains},\, \texttt{realizes},\, \texttt{specifies},\, \texttt{depends\_on},\, \texttt{conforms\_to}, \\
                  &\texttt{verified\_by},\, \texttt{evidenced\_by},\, \texttt{constrained\_by},\, \texttt{supersedes},\, \texttt{affects}\}.
\end{align*}

\begin{definition}[Provenance-Typed Relation]
\label{def:relation}
A relation $r = (u, w, \tau, o, c, q, \nu)$ connects valid endpoint types $(u, w) \in \mathcal{V} \times \mathcal{V}$ ($u \neq w$) with relation label $\tau \in \mathcal{R}$, origin record $o = (\texttt{type}, \texttt{source})$ where $o.\texttt{type} \in \{\texttt{explicit}, \texttt{static}, \texttt{inferred}\}$ and $o.\texttt{source} \neq \emptyset$, confidence $c \in [0, 1]$, boolean verification flag $q \in \{\texttt{true}, \texttt{false}\}$, and verification tuple $\nu = (\texttt{verifier}, \texttt{method}, \texttt{timestamp}, \texttt{version})$ required if and only if $q = \texttt{true}$.
\end{definition}

\begin{definition}[Four-Dimensional Epistemic Entity State]
\label{def:epistemic_state}
For any governed engineering entity $e$, EPE maintains an orthogonal state tuple $\mathcal{S}(e) = (d, m, r, v)$ across four independent domains:
\begin{align}
  d &\in \{\texttt{draft},\, \texttt{reviewed},\, \texttt{approved}\}, \quad
  m \in \{\texttt{not\_started},\, \texttt{in\_progress},\, \texttt{implemented}\}, \nonumber\\
  r &\in \{\texttt{pending},\, \texttt{passed},\, \texttt{failed},\, \texttt{error},\, \texttt{not\_run}\}, \quad
  v \in \{\texttt{valid},\, \texttt{stale},\, \texttt{invalid},\, \texttt{unknown}\}.
\end{align}
\end{definition}

Decoupling $\mathcal{S}(e)$ prevents state masking across agent turns: combinations such as $(\texttt{approved},\allowbreak \texttt{implemented},\allowbreak \texttt{passed},\allowbreak \texttt{stale})$ or $(\texttt{approved},\allowbreak \texttt{implemented},\allowbreak \texttt{failed},\allowbreak \texttt{valid})$ remain first-class representable states. Operationalizing Assumption-Based Truth Maintenance Systems (ATMS)~\cite{dekleer1986atms} and Pollock-style defeasible reasoning~\cite{pollock1987defeasible}, historical approvals ($d$) and verification outcomes ($r$) act as defeasible justifications grounded in an explicit working-tree assumption slice $\mathcal{A}(E_0, G) \subseteq \text{dom}(\mathcal{M}(\mathcal{W}))$. Building on the four coarse disk states, EPE computes a five-valued defeasible evidence assessment $v_{\text{def}} \in \{\texttt{valid}, \texttt{undercut}, \texttt{rebutted}, \texttt{invalid}, \texttt{unknown}\}$: mutating any assumption in $\mathcal{A}(E_0, G)$, changing verification policy, or exceeding \texttt{max\_age\_seconds} \emph{undercuts} prior passing evidence, whereas a failing test assertion under intact assumptions \emph{rebuts} it. Computing a minimal sufficient test subset over $\mathcal{W}$ is an NP-hard set-cover problem; EPE avoids that intractability by validating the configured policy check set $\mathcal{P}_{\text{verify}}.\texttt{checks}$ against the sound ATMS dependency slice $\mathcal{A}(E_0, G)$. Furthermore, BTAE accommodates heterogeneous maintenance tasks without forcing waterfall documentation by routing four workflow types through schema-enforced mandatory task fields before activation: \texttt{feature} (\texttt{objective}, \texttt{acceptance}, \texttt{target\_entities}), \texttt{bugfix} (plus \texttt{reproduction}), \texttt{refactor} (plus \texttt{baseline\_manifest}), and \texttt{exploration} (plus \texttt{hypothesis}, where prototypes remain isolated from production design approval unless promoted under the derived risk floor).

\subsection{Deterministic Atomic Graph Construction and Incremental Maintenance}
\label{sec:method_graph}

Maintaining an engineering knowledge graph inside an active repository faces two systems hazards: external graph databases introduce state drift away from Git working trees, while in-place JSON index updates risk exposing partial snapshots during concurrent reads, crashes, or mid-build file edits.

EPE constructs the EKG purely from authoritative repository files and publishes content-addressed generations atomically. Given repository root $\mathcal{W}$ and configuration $\mathcal{C}$, EPE first computes a sorted SHA-256 input manifest $\mathcal{M}(\mathcal{W}) = \big((p_1, h_1), \dots, (p_k, h_k)\big)$ with canonical JSON digest $H(\mathcal{M})$, rejecting any symlinked root, symlinked directory, or path escaping $\mathcal{W}$, while excluding derived indices (\texttt{.engineering/}) and reports (\texttt{reports/}) so inspection never self-invalidates inputs. Next, the document parser extracts frontmatter entities and explicit relations, while the Python AST analyzer extracts qualified class/function symbols, signatures, and static imports across managed \texttt{code\_roots}. Each static import is resolved deterministically into one of three edge classes: (i) a resolved internal dependency (\texttt{confidence} $= 1.0$), (ii) an external package dependency \texttt{EXT-*} (\texttt{confidence} $= 0.5$), or (iii) an ambiguous/unresolvable target \texttt{UNKNOWN-*} (\texttt{confidence} $= 0.0$) accompanied by warning \texttt{EPE-AST-IMPORT-UNKNOWN}; dynamic constructs such as \texttt{importlib.import\_module} and \texttt{\_\_import\_\_} emit explicit \texttt{EPE-AST-DYNAMIC} blind-spot diagnostics. When \texttt{module\_roots} maps modules \texttt{MOD-*} to directory prefixes, EPE verifies unique file ownership and lifts cross-module file imports into module-level \texttt{depends\_on} edges.

To publish without races, EPE deduplicates identical edges by canonical SHA-256 digest, sorts nodes and edges deterministically (excluding wall-clock timestamps from the payload), re-reads $\mathcal{M}'(\mathcal{W})$ to abort with \texttt{EPE-GRAPH-INPUT-CHANGED} if any file mutated during parsing, and computes snapshot digest $g = \text{SHA256}(\text{payload})$. Under an exclusive POSIX file lock (\texttt{.write.lock}), EPE writes \texttt{entities.json} and \texttt{graph.json} into temporary directory \texttt{.tmp-}$g$, atomically renames it to \texttt{.engineering/generations/}$g$, updates \texttt{current.json} as the single commit pointer, seeds a Directory-Merkle stat cache (\texttt{manifest\_stat\_cache.json}), and exposes bounded generation retention (\texttt{gc\_generations}) preserving \texttt{current.json} and the $k-1$ newest snapshots.

For fast iterative edits, \texttt{build\_graph\_incremental} validates the prior snapshot's digest, parser configuration, and stored \texttt{CodeParseResult} records, computes the changed path set $\Delta = \{p \in \mathcal{M}_{\text{old}} \cup \mathcal{M}_{\text{curr}} \mid h_{\text{old}}(p) \neq h_{\text{curr}}(p)\}$ via a Git-style \emph{50\,ms racy-clean stat-delta manifest} (\texttt{build\_manifest\_cached}), re-parses only modified Python files in $\Delta$, and merges them with verified cached facts---falling back to a full AST parse whenever schema or parser versions differ. Any file whose filesystem inode, byte size, \texttt{mtime\_ns}, or \texttt{ctime\_ns} changed, or whose \texttt{mtime\_ns} lies within $\Delta_{\text{racy}} = 50\,\text{ms}$ of cache persistence, is conservatively re-hashed from disk; meanwhile, \texttt{build\_merkle\_tree} arranges entries into a deterministic directory hierarchy supporting $O(\log |\mathcal{W}|)$ inclusion proofs (\texttt{merkle\_proof}, \texttt{verify\_merkle\_proof}).

\begin{proposition}[Deterministic Content Equivalence]
\label{prop:incremental}
Let $G_{\text{prev}}$ be a complete, digest-verified snapshot under parser configuration $\pi$, and let $\mathcal{W}'$ be any repository state obtained by arbitrary file additions, modifications, or deletions under identical $\pi$. If no concurrent write mutates $\mathcal{W}'$ during construction, \texttt{build\_graph\_incremental} produces the exact same canonical node list, edge list, and SHA-256 \texttt{content\_digest} as a cold \texttt{build\_graph} execution over $\mathcal{W}'$.
\end{proposition}

Because timestamps are excluded from canonical hashing, racy-window files are conservatively re-hashed, and cached AST facts are validated against per-file SHA-256 digests before reuse, Proposition~\ref{prop:incremental} guarantees zero incremental drift while eliminating external database infrastructure. In asymptotic time complexity, cold \texttt{build\_graph} runs in $O(\sum_{p \in \mathcal{W}} |\text{bytes}(p)| + |\mathcal{V}| + |\mathcal{E}|\log|\mathcal{E}|)$. Without stat filtering, full-tree manifest hashing incurs an unavoidable $O(\sum_{p \in \mathcal{W}} |\text{bytes}(p)|)$ term even when $|\Delta|=1$; with the racy-safe stat-delta manifest (\texttt{build\_manifest\_cached}), incremental rebuilds reduce byte reading and AST parsing to $O(|\mathcal{W}|\,T_{\text{stat}} + \sum_{p \in \Delta_{\text{stat}}} |\text{bytes}(p)|) + O(|\mathcal{V}| + |\mathcal{E}|\log|\mathcal{E}|)$ while preserving global import resolution, module aggregation, edge deduplication, canonical sorting, and \texttt{fsync}-backed atomic publication.

\subsection{Stratified Bounded Context Assembly and Bidirectional Impact Analysis}
\label{sec:method_context}

Passing an entire repository to a coding agent wastes context budget and dilutes task-critical constraints, whereas pure embedding similarity retrieval frequently omits mandatory governance rules or drops downstream consumers affected by an interface change.

Given a task $T$, fresh snapshot $G$, and item budget $B$, \texttt{resolve\_context} assembles a provenance-tagged bundle prioritized across four epistemic strata: \texttt{C0} mandatory charter and rules, \texttt{C1} target and graph-adjacent designs, \texttt{C2} mapped code, tests, and lexical matches, and \texttt{C3} historical decisions and task evidence; full strata definitions appear in Appendix~\ref{app:strata}.
Each item is capped at 4,000 characters with explicit \texttt{truncated\_paths} and budget \texttt{omissions} tracking. Crucially, \texttt{store\_cached\_context} enforces a \emph{fail-closed caching rule}: any bundle with \texttt{incomplete = true} (due to omissions, item truncation, or mid-assembly file mutation) is never persisted to cache, indexes the entry by \texttt{context\_cache\_key} (v6, hashing the selection-policy version, budget $B$, task fields, required-source fingerprints, source assessments, snapshot \texttt{content\_digest}, and full \texttt{context\_inputs\_manifest} digest including \texttt{engineering/epe.yaml}), and re-verifies every cached item's SHA-256 byte digest on read.

\begin{lemma}[Fail-Closed Context Cache Freshness]
\label{lem:cache_freshness}
Under SHA-256 collision resistance, if \texttt{load\_cached\_context} returns a cached bundle $B_c$ over workspace $\mathcal{W}$ for selection parameters $(T, G, B, \mathcal{C})$, then (i) $B_c$ satisfied $\texttt{incomplete} = \texttt{false}$ (zero item truncations and zero budget omissions) when persisted, (ii) the selection-policy key (\texttt{version}, $B$, $T$, \texttt{required\_sources}, $G.\texttt{content\_digest}$) matches identically, and (iii) both the composite context-inputs manifest (including \texttt{engineering/epe.yaml}, policies, approvals, audit logs, and evidence) and every cached item's byte digest match $\mathcal{W}$ identically (proof in Appendix~\ref{app:proof}).
\end{lemma}

For change impact analysis, unidirectional graph reachability is unsound because \texttt{realizes} points from design to code whereas \texttt{depends\_on} points from consumer to provider. Given seed entities $E_0$ and node budget $B_{\text{imp}}$, \texttt{analyze\_impact} constructs a direction-aware propagation graph traversing \emph{forward} along $\{\texttt{realizes}, \texttt{verified\_by}, \texttt{evidenced\_by}, \texttt{conforms\_to}, \texttt{specifies}, \texttt{affects}\}$, \emph{reverse} along $\{\texttt{realizes}, \texttt{depends\_on}\}$ (so modifying a provider propagates to its callers and owning designs), and bidirectionally between \texttt{Source} files and their AST-defined \texttt{Symbol} nodes. Each reached node $u$ is classified as \texttt{determined} if and only if at least one propagation path from $E_0$ to $u$ consists exclusively of \texttt{explicit} or \texttt{static} edges with $\texttt{confidence} = 1.0$; otherwise $u$ is labeled \texttt{candidate}, and a priority-upgrade rule re-propagates downstream if a later \texttt{determined} path reaches a previously \texttt{candidate} node. Any AST blind spot in $G$ automatically marks the impact report \texttt{incomplete = true}.

\begin{lemma}[Termination and Complexity of Priority-Upgrading Impact Propagation]
\label{lem:impact_termination}
For any finite snapshot $G = (\mathcal{V}, \mathcal{E})$ and node budget $B_{\text{imp}}$, \texttt{analyze\_impact} upgrades each reached node's certainty label monotonically along $\texttt{candidate} \prec \texttt{determined}$ at most once, terminating in $O(|\mathcal{V}| + |\mathcal{E}|)$ edge inspections while visiting at most $B_{\text{imp}}$ distinct nodes (proof in Appendix~\ref{app:proof}).
\end{lemma}

\subsection{Fail-Closed Governance Gates, Cryptographic Approvals, and Host Boundaries}
\label{sec:method_governance}

Even with an accurate knowledge graph, autonomous agents can bypass prompt-level instructions by editing frontmatter (\texttt{design\_status:\ approved}), claiming tests passed without running them, modifying files outside their assigned task, or assuming an unverified execution host enforces filesystem sandboxing.

The full pseudocode of EPE's composite completion gate (\texttt{epe check}) is given in Algorithm~\ref{alg:finalize} (Appendix~\ref{app:gate_algo}); the gate enforces five orthogonal invariants:
\begin{itemize}[leftmargin=1.6em,itemsep=2pt,topsep=3pt]
  \item \textbf{Revision-Bound SSH Approvals and Derived Risk Floors (\texttt{G1}):} EPE computes a canonical design revision $H_{\text{rev}}(e)$ (\texttt{epe-design-revision-v3}) hashing $e$'s normalized declaration, raw Markdown/YAML bytes, and the byte digests of all referenced \texttt{CONTRACT-*} files (whose Draft 2020-12 schemas are checked with external network \texttt{\$ref} URIs rejected fail-closed). Because approvals are stored externally under \texttt{engineering/approvals/*.json} rather than inside the document itself, signing $H_{\text{rev}}(e)$ incurs zero frontmatter circularity: self-edited \texttt{design\_status:\ approved} frontmatter never satisfies \texttt{assess\_approval}. To eliminate task-level self-downgrade, \texttt{derive\_risk\_floor} computes a structural minimum risk floor $\rho_{\text{floor}}(T, \Delta\mathcal{W}, G) \in \{\texttt{low}, \texttt{medium}, \texttt{high}\}$ from target entity types (\texttt{Architecture}/\texttt{Contract} $\to \texttt{high}$; \texttt{Module} or multi-target edits $\to \texttt{medium}$), modified \texttt{high\_risk\_paths}, \texttt{determined} EKG blast-radius thresholds, and incomplete-graph fail-closure. Under \texttt{enforce\_risk\_floor = true}, EPE sets $r_{\text{eff}}(T) = \max(T.\texttt{risk}, \rho_{\text{floor}})$, emits \texttt{EPE-RISK-DOWNGRADE} whenever $T.\texttt{risk} < \rho_{\text{floor}}$, and enforces $r_{\text{eff}}(T)$ against OpenSSH Ed25519 signatures (\texttt{ssh-keygen -Y verify}, namespace \texttt{epe-approval}) committed in \texttt{events.jsonl} (protected by \texttt{O\_NOFOLLOW}, \texttt{flock}, and \texttt{fsync}); medium-risk approvals disallow ChangeSet self-approval, and high-risk approvals require $\texttt{kind} = \texttt{human}$.
  \item \textbf{AST Mapping and ADR-Scoped Architecture Fitness (\texttt{G2}--\texttt{G3}):} Mapping validation blocks any \texttt{implemented} entity whose target file or qualified AST symbol is missing. Architectural validation checks dependency cycles (\texttt{ARCH-CYCLE}), package \texttt{allowed\_imports}, forbidden dependency globs, and AST blind spots (blocking when \texttt{unknown\_is\_blocking = true}, the policy default). Policy exceptions waive a forbidden edge if and only if they reference an entity $\texttt{ADR-*} \in \mathcal{E}$ with $\texttt{design\_status} = \texttt{approved}$, a valid future UTC \texttt{review\_until} deadline, and matching $(\texttt{rule\_id}, \texttt{from}, \texttt{to})$ edge globs.
  \item \textbf{Manifest-Bound Subprocess Evidence and ATMS Slicing (\texttt{G4}):} \texttt{run\_checks} executes configured \texttt{argv} arrays without shell interpolation, enforces process-group timeouts, redacts secret tokens from logs, parses \texttt{pytest} JUnit XML reports (rejecting zero-test runs as \texttt{test\_execution\_unconfirmed}), snapshots before/after input manifests over $\mathcal{M}_{\text{ver}}(\mathcal{W})$ (covering \texttt{design\_roots}, \texttt{code\_roots}, \texttt{test\_roots}, \texttt{contract\_roots}, \texttt{engineering/epe.yaml}, and \texttt{engineering/policies}), and atomically publishes immutable \texttt{RUN-*} bundles. At check time, \texttt{finalize\_gate} enforces strict timezone-aware UTC ordering over \texttt{finished\_at} (\texttt{EPE-GATE-EVIDENCE-ORDER}), and \texttt{evidence\_validity} recomputes $v \in \{\texttt{valid}, \texttt{stale}, \texttt{invalid}, \texttt{unknown}\}$ from scratch against current disk artifacts, policy hash, Python/OS tuple, optional \texttt{max\_age\_seconds}, and $\mathcal{M}_{\text{ver}}(\mathcal{W})$. Companion routine \texttt{assess\_defeasible\_validity} computes the sound upstream ATMS assumption slice $\mathcal{A}(E_0, G)$ (failing closed to full $\mathcal{M}_{\text{ver}}(\mathcal{W})$ on any AST blind spot or missing entity) and classifies failures into \texttt{undercut} (policy, temporal, or assumption defeat) versus \texttt{rebutted} (failing execution under intact assumptions).
  \item \textbf{ChangeSet Baselines, Leases, and Host Capability Attestation (\texttt{G5}):} Before editing, \texttt{capture\_baseline} records SHA-256 digests of all workspace files and symlinks; at completion, \texttt{compare\_baseline} verifies that every modified path satisfies \texttt{write\_scope} while respecting \texttt{effective\_protected\_paths} (the union of ChangeSet \texttt{protected} globs and non-downgradeable \texttt{governance\_policy.protected\_paths} in \texttt{epe.yaml}). For multi-agent coordination, \texttt{detect\_changeset\_conflicts} conservatively flags overlapping \texttt{planned}/\texttt{active} globs, and \texttt{acquire\_lease} manages TTL-bounded exclusive POSIX leases with automatic post-expiry crash recovery. Finally, EPE separates advisory planning from host trust: \texttt{btae-mcp} exposes strictly read-only MCP tools (\texttt{btae\_project\_context}, \texttt{btae\_design\_plan}, \texttt{btae\_validate\_project}), \texttt{local-cli} and \texttt{HermesACPAdapter} default to \texttt{handshaken-untrusted}, and \texttt{HermesHostAdapter} grants \texttt{handshaken} status only when the host signs $\{\texttt{protocol\_version}, \texttt{nonce}, \texttt{capabilities}\}$ with a trusted Ed25519 key within its configured operation subset.
\end{itemize}

\begin{lemma}[Evidentiary Validity Monotonicity]
\label{lem:validity_monotonicity}
Let $E^*$ be an evidence bundle with $v(E^*, \mathcal{W}) = \texttt{valid}$ bound to input manifest $\mathcal{M}_{\text{ver}}(\mathcal{W})$ (covering \texttt{design\_roots}, \texttt{code\_roots}, \texttt{test\_roots}, \texttt{contract\_roots}, \texttt{engineering/epe.yaml}, and \texttt{engineering/policies}), verification policy digest $h_{\text{pol}}$, and artifact digests $\{h_a\}$. Under SHA-256 collision resistance, any single-byte modification, addition, or deletion in any path $p \in \text{dom}(\mathcal{M}_{\text{ver}}(\mathcal{W})) \cup \text{artifacts}(E^*) \cup \{\text{policy}(\mathcal{C})\}$ yields $v(E^*, \mathcal{W}') \neq \texttt{valid}$ (i.e., $v(E^*, \mathcal{W}') \in \{\texttt{stale}, \texttt{invalid}, \texttt{unknown}\}$), and under complete ATMS slice $\mathcal{A}(E_0, G)$ any mutation in $\mathcal{A}(E_0, G)$ yields $v_{\text{def}} = \texttt{undercut}$ (proof in Appendix~\ref{app:proof}).
\end{lemma}

\begin{lemma}[Composite Gate Soundness under Anchored Policy Boundary]
\label{lem:gate_soundness}
Suppose (i) policy-boundary inputs (\texttt{engineering/epe.yaml}, \texttt{engineering/policies/*}, and EPE engine modules) are integrity-anchored (e.g., via CI branch protection or read-only host mounts), (ii) all behavior-affecting workspace paths lie inside $\text{dom}(\mathcal{M}_{\text{ver}}(\mathcal{W}))$ and Python \texttt{code\_roots}, and (iii) \texttt{unknown\_is\_blocking = true}. Then for any adversarial instance across the twelve in-coverage fault families---approval forgery, approval staleness, mapping drift, ADR lifecycle violations, evidence faults, scope escapes, audit tampering, host attestation violations, static forbidden imports, direct dynamic imports, task risk downgrades, and local policy/engine tampering---Algorithm~\ref{alg:finalize} returns \texttt{blocked} or \texttt{error} (proof in Appendix~\ref{app:proof}).
\end{lemma}

\paragraph{Threat Model and Policy-Boundary Trust.}
EPE treats the autonomous coding agent as an untrusted writer of source code, Markdown frontmatter, task risk declarations, ChangeSet \texttt{protected} lists, and local completion claims: under \texttt{enforce\_risk\_floor = true}, unilateral attempts to downgrade \texttt{T.risk} or omit ChangeSet \texttt{protected} globs are blocked locally via \texttt{derive\_risk\_floor} and \texttt{effective\_protected\_paths}. Remaining \emph{policy-boundary inputs} are project configuration (\texttt{engineering/epe.yaml}, including \texttt{enforce\_risk\_floor}, trust roots, and \texttt{unknown\_is\_blocking}), policy files, and the EPE engine modules themselves: if an unisolated local agent holds write permissions to disable policy checks or stub engine modules on disk, local execution alone cannot prevent self-tampering. Mandatory enforcement against uncooperative local processes therefore pairs EPE with protected-branch CI gates, external secret-managed signing keys, or read-only policy mounts under \texttt{HostPort v1}. In multi-turn deployments, blast-radius-derived risk floors also surface a potential \emph{governance arbitrage} (``salami-slicing'') incentive where an agent splits edits into small low-risk patches; EPE mitigates this via cumulative ChangeSet baselines and structural entity floors.

\section{Experimental Evaluation}
\label{sec:experiments}

We evaluate BTAE and the Engineering Protocol Engine (EPE) across five research questions designed to test determinism, governance safety and liveness, context efficiency, scalability, and coordination boundaries:
\begin{itemize}[leftmargin=1.6em,itemsep=2pt,topsep=3pt]
  \item \textbf{RQ1 (Determinism \& Interactive Latency):} Does EPE reconstruct byte-identical knowledge graphs from scratch and execute governance checks at interactive CLI latencies on a real self-hosted codebase?
  \item \textbf{RQ2 (Governance Safety, Liveness \& Ablation):} How effectively does EPE's four-dimensional epistemic gate prevent false acceptances and false rejections compared with self-reported, static-schema, and boolean-execution baselines?
  \item \textbf{RQ3 (Bounded Context Compression \& Freshness):} How much context payload does stratified \texttt{C0}--\texttt{C3} resolution save relative to full-repository loading, and does content-addressed caching invalidate reliably on single-byte edits?
  \item \textbf{RQ4 (Impact Blast Radius \& Incremental Graph Equivalence):} Does direction-aware impact propagation expose architectural layer risk, and does incremental graph construction preserve exact digest equivalence as repositories scale?
  \item \textbf{RQ5 (Audit Integrity, Leases \& Host Boundaries):} What is the overhead of append-only SHA-256 hash-chain auditing, POSIX ChangeSet leases, and conservative multi-agent conflict detection?
\end{itemize}

\subsection{Experimental Setup and Self-Hosted Benchmark}
\label{sec:exp_setup}

All experiments execute on Linux 7.0.0-43-generic (\texttt{x86\_64}, \texttt{glibc 2.43}) running on a 20-core 13th-Gen Intel Core i5-13600K host with 30.5\,GB RAM, CPython 3.14.4, and \texttt{pytest} 9.1.1 against isolated copies of the self-hosted EPE repository (commit \texttt{d67f9295}), a 7,338-LOC reference implementation whose full composition is detailed in Appendix~\ref{app:repo}. Running the complete automated verification suite executes \textbf{368 tests in 56.92\,s} (57.4\,s subprocess wall clock) with exit code 0, corroborated by remote GitHub Actions CI and protected-branch enforcement on \texttt{master}.

To isolate the causal contribution of each BTAE layer and policy-boundary assumption in RQ2, we compare five baselines---\textbf{B0} (self-reported, prompt-only), \textbf{B1} (static AST and schema only), \textbf{B2} (boolean execution), \textbf{B\textsubscript{git}} (minimal Git \texttt{write-tree} + signed tag + diff scope), and \textbf{B3} (strong practical stack combining \texttt{B1} $\cup$ \texttt{B2} $\cup$ \texttt{B\textsubscript{git}} + CODEOWNERS + CI branch protection)---against three deployment modes of \textbf{BTAE / EPE}: unanchored declared risk (\textbf{EPE\textsubscript{unanch}}), local policy-derived risk floors (\textbf{EPE\textsubscript{floor}}), and anchored CI/policy-boundary governance (\textbf{EPE\textsubscript{full}}); exact definitions appear in Appendix~\ref{app:baselines}.

\subsection{RQ1: Self-Hosted Determinism and Operation Latencies}
\label{sec:exp_rq1}

On the self-hosted repository, \texttt{epe graph build} publishes a complete Engineering Knowledge Graph of \textbf{426 nodes} and \textbf{550 provenance-typed edges} (\texttt{graph.json}: 519.67\,KB; \texttt{entities.json}: 9.03\,KB; \texttt{manifest\_stat\_cache.json}: 29.60\,KB), with node- and edge-type decompositions in Appendix~\ref{app:repo}. The managed input manifest $\mathcal{M}_{\text{graph}}(\mathcal{W})$ covers 187 of 341 non-derived repository files (54.84\% of files, 1,954,409 of 2,270,718 bytes = \textbf{86.07\%} of repository bytes), leaving benchmark, scratch, and tooling scripts outside governed roots. Architectural validation inspects all 430 import statements across the 9 logical modules, confirming 0 dependency cycles, 0 forbidden-boundary violations, and 0 unresolvable AST blind spots. Across 10 consecutive cold rebuilds in which \texttt{.engineering/} was deleted prior to each run, EPE produced \textbf{1 unique SHA-256 \texttt{content\_digest}}, empirically confirming deterministic snapshot reconstruction (\texttt{KR-03} / \texttt{AC-10}).

Table~\ref{tab:rq1_latency} in Appendix~\ref{app:latency} separates \emph{advisory interactive} operations from \emph{governance gate/batch} execution across $N = 25$ trials. Every advisory check executes well below human or LLM turn latency: policy validation takes 2.04\,ms (p50), impact propagation takes 4.28\,ms, and cold \texttt{graph build} across all 54 managed Python source files completes in 132.18\,ms (p50). Rebuilding incrementally after a single-file modification with the 50\,ms racy-safe stat-delta manifest cuts median build latency to \textbf{57.51\,ms} ($2.30\times$ median speedup), while still executing global import resolution, edge deduplication, canonical sorting, and \texttt{fsync}-backed atomic publication (causing occasional flush tail spikes up to 162.98\,ms vs.\ cold max 138.90\,ms). SHA-256 verified context-cache hits execute in \textbf{1.98\,ms} ($7.75\times$ faster than cold assembly) when measured on a complete non-truncated bundle; on the self-hosted repository itself, \texttt{AGENTS.md} (9.5\,KB) exceeds the 4,000-character per-item cap, so EPE marks the bundle \texttt{incomplete = true} and conservatively refuses to persist a truncated cache entry (\texttt{self\_hosted\_truncated\_cache\_refused = true}). By contrast, governance completion budgets reflect verification cost: gate-only \texttt{epe check} over fresh pre-recorded evidence runs in \textbf{57.83\,ms} (p50), an end-to-end mini-suite \texttt{verify + check} batch runs in \textbf{548.75\,ms} (p50), and the full 368-test self-hosted suite takes \textbf{56.92\,s}.

\subsection{RQ2: Governance Gate Safety, Liveness, and Ablation}
\label{sec:exp_rq2}

\begin{figure}[t]
  \centering
  \includegraphics[width=\textwidth]{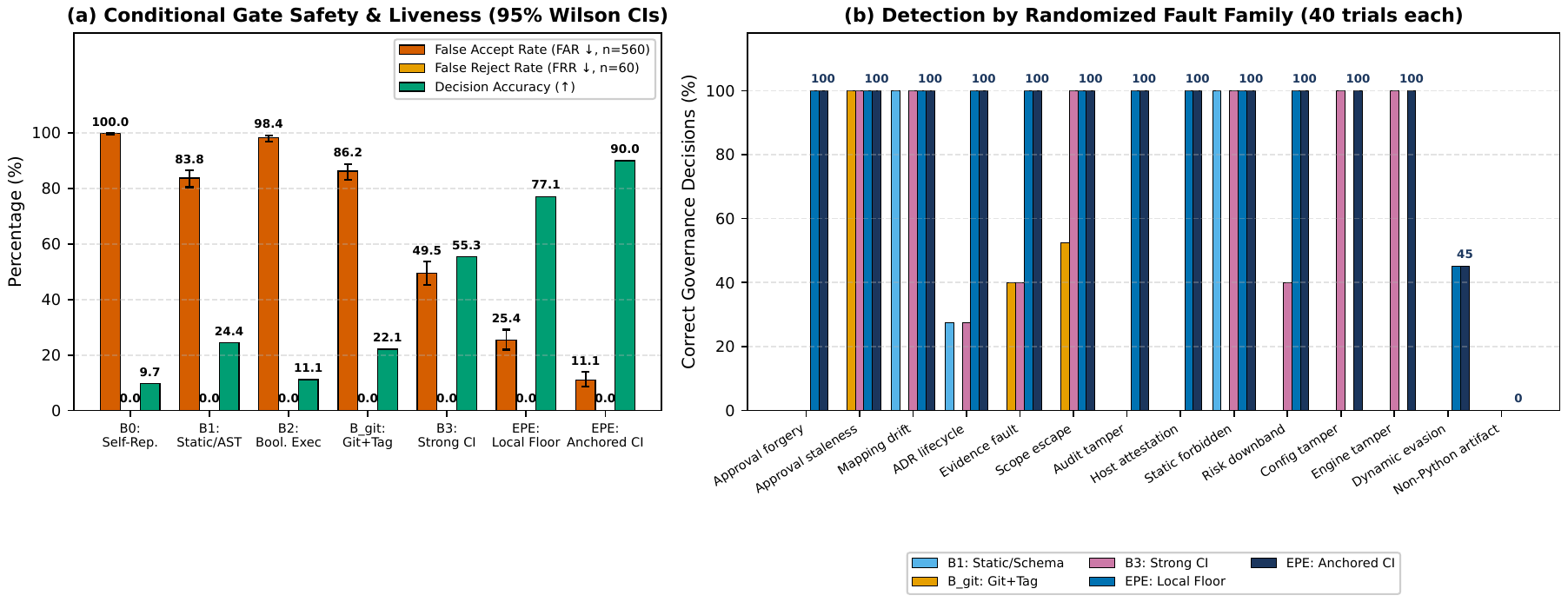}
  \caption{Conditional governance ablation over 560 randomized adversarial trials (14 fault families) and 60 valid workflows. (a) False Acceptance Rate (FAR) with 95\% Wilson CIs, False Rejection Rate (FRR), and accuracy across \textbf{B0}--\textbf{B3}, local risk-floor EPE, and CI-anchored EPE. (b) Correct-decision rate by fault family (40 trials each), isolating policy-boundary tampering (\texttt{config-tamper}, \texttt{engine-tamper}) and the two out-of-coverage AST/polyglot boundary families.}
  \label{fig:gate_ablation}
\end{figure}

We evaluate all eight governance regimes/modes over a seeded, randomized fault-injection suite: 14 adversarial fault families (40 trials each, $N = 560$) spanning the original 11 families (approval forgery and staleness, mapping drift, ADR waiver lifecycle, evidence validity, ChangeSet scope, audit-chain integrity, host capability attestation, statically forbidden imports, plus the two out-of-coverage boundary families---dynamic-import evasion and non-Python artifacts) and three new policy-boundary red-team families (\texttt{risk-downband}, \texttt{config-tamper}, and \texttt{engine-tamper}), alongside 60 randomized valid workflows. All fault parameters are drawn from a fixed-seed RNG (`20260520`) and judged on identical repository states, yielding exact paired measurements whose Wilson intervals quantify trial sampling rather than engine stochasticity. As summarized in Figure~\ref{fig:gate_ablation} and Table~\ref{tab:rq2_ablation} (Appendix~\ref{app:ablation}), over all 560 adversarial trials \textbf{B0} accepts 100.00\% $[99.32, 100.00]$, \textbf{B2} reaches 98.39\% FAR, \textbf{B1} reaches 83.75\% FAR, and minimal Git tree-hash/tag checking (\textbf{B\textsubscript{git}}) reaches \textbf{86.25\% FAR} $[83.15, 88.86]$ (blocking post-tag tree edits, failed exits, and out-of-scope paths while missing AST mapping, ADR expiry, evidence metadata/age/zero-test checks, audit chains, host attestation, and risk downgrades). Combining them into the strong practical \textbf{B3} stack cuts FAR to \textbf{49.46\%} $[45.34, 53.59]$ (277/560 false accepts; 57.50\% on the original 11 families).

Conditional deployment measurements quantify the exact impact of policy-boundary protections: without derived risk floors, \textbf{EPE\textsubscript{unanch}} suffers \textbf{32.50\% FAR} $[28.75, 36.49]$ (182/560, missing all 120 trials across \texttt{risk-downband}, \texttt{config-tamper}, and \texttt{engine-tamper}); enabling local policy-derived risk floors (\textbf{EPE\textsubscript{floor}}) eliminates all 40/40 \texttt{risk-downband} attacks and lowers local FAR to \textbf{25.36\%} $[21.93, 29.12]$ (142/560); and anchoring policy/engine integrity under CI or read-only mounts (\textbf{EPE\textsubscript{full}}) reaches \textbf{11.07\% FAR} $[8.73, 13.94]$ (62/560; \textbf{90.00\%} accuracy, \textbf{0/60 FRR} on standard compliant workflows; 14.09\% FAR $[11.15, 17.65]$ on the original 11-family $N=440$ subset). Because narrower baselines target subsets of faults, Figure~\ref{fig:gate_ablation}(b) and Table~\ref{tab:scenarios_summary} provide the fair per-family breakdown. Every silent acceptance under \textbf{EPE\textsubscript{full}} lies strictly in the two documented boundary families: all 40 non-Python artifacts and 22 of 40 reflective dynamic-loading forms (\texttt{importlib.util.find\_spec}, aliased \texttt{import\_module}, \texttt{sys.modules}), while 18 of 40 direct dynamic forms remain blocked as \texttt{EPE-AST-DYNAMIC} and zero silent acceptances occur across the 12 in-coverage families (0/480; rule-of-three upper bound 0.63\%). Paired McNemar tests confirm \textbf{EPE\textsubscript{full}} outperforms every comparator ($p < 10^{-6}$): 215/0 discordant blocks vs.\ \textbf{B3}, 421/0 vs.\ \textbf{B\textsubscript{git}}, 407/0 vs.\ \textbf{B1}, 489/0 vs.\ \textbf{B2}, and 498/0 vs.\ \textbf{B0}. Leave-one-out gate attribution (Table~\ref{tab:rq2_ablation}) confirms non-redundant defense-in-depth across all five gates: disabling \texttt{G1} raises FAR by $+23.93$\,pp (134 unique blocks), \texttt{G5} by $+22.32$\,pp (125 unique), \texttt{G3} by $+21.07$\,pp (118 unique), \texttt{G4} by $+9.11$\,pp (51 unique), and \texttt{G2} by $+7.14$\,pp (40 unique). Conversely, the 0/60 FRR figure holds only on standard compliant workflows: probing 15 valid tasks using standard-library lazy loading (\texttt{importlib.import\_module('json')}) yields 15/15 conservative blocks (100\% FRR on that boundary class) under \texttt{unknown\_is\_blocking = true}.

\subsection{RQ3: Bounded Context Compression and Cache Freshness}
\label{sec:exp_rq3}

\begin{figure}[t]
  \centering
  \includegraphics[width=\textwidth]{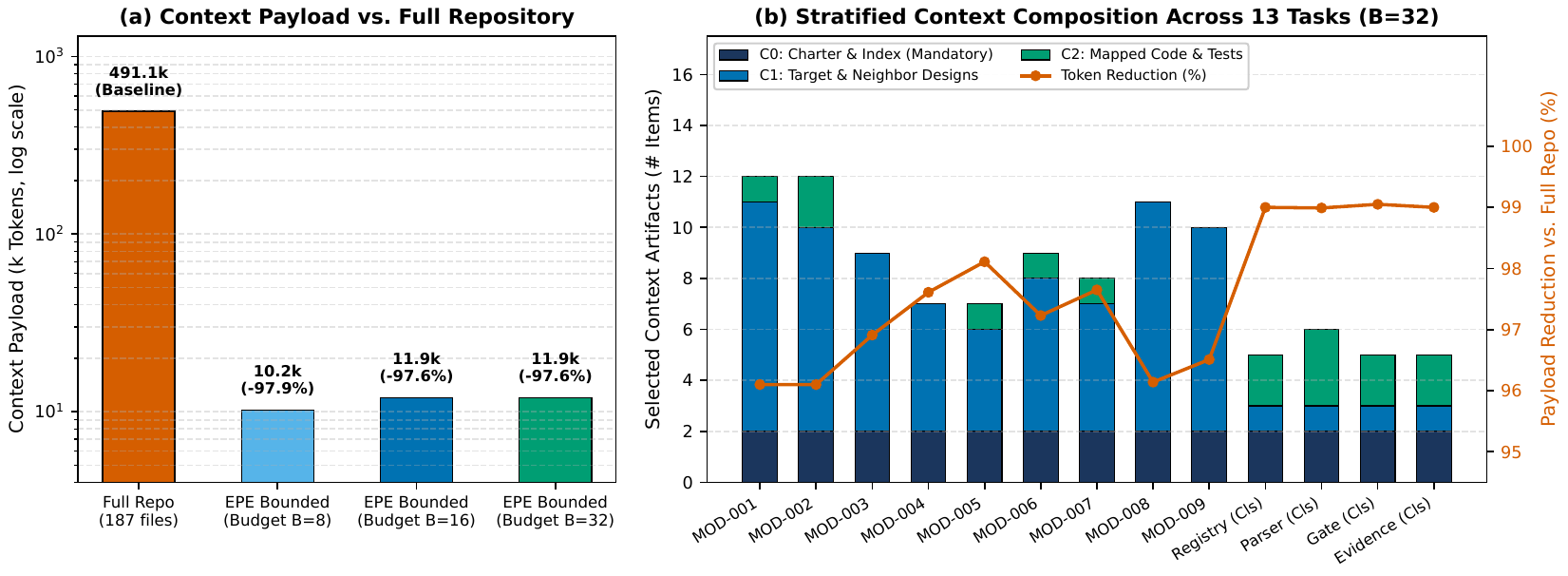}
  \caption{Stratified context assembly (\texttt{C0}--\texttt{C3}) evaluation across 13 maintenance tasks. (a) Log-scale token comparison against full-repository context loading. (b) Per-task artifact breakdown across strata \texttt{C0}, \texttt{C1}, and \texttt{C2} alongside token reduction percentage at $B = 32$.}
  \label{fig:context_compression}
\end{figure}

Loading all 190 managed design, contract, source, policy, and test files in the self-hosted repository consumes 1,964,293\,B ($\approx 491.1\text{k}$ tokens under a 4-byte/token heuristic), which serves as a full-repository upper-bound reference rather than an interactive retrieval baseline such as grep or BM25. We evaluate \texttt{resolve\_context} across 13 representative maintenance tasks spanning all four workflow types, targeting each of the 9 modules and 4 core classes. As shown in Figure~\ref{fig:context_compression}, with the full budget breakdown in Table~\ref{tab:rq3_context} (Appendix~\ref{app:exp_tables}), EPE reduces mean context payload by \textbf{97.92\%} at budget $B = 8$ (10.2k tokens) and \textbf{97.57\%} at $B \in \{16, 32\}$ (11.9k tokens; both budgets yield identical 8.15-item averages because the self-hosted 1-hop neighborhood saturates below 16 items), while preserving mandatory \texttt{C0} charter retention and target-design recall on all 13 of 13 tasks across all budgets. Under tight budget $B = 8$, EPE never drops \texttt{C0} silently; instead, it explicitly logs an average of 6.23 omitted neighbor paths in \texttt{omissions[]}, which drops to zero at $B \ge 16$. Fine-grained class tasks achieve even stronger compression (\textbf{98.99\%--99.05\%}) than broad module tasks (\textbf{96.10\%--98.11\%}) because class entities bind directly to specific source and test files. Exercising Lemma~\ref{lem:cache_freshness} empirically across 8 single-byte mutation probes spanning \texttt{AGENTS.md}, designs, Python sources, tests, and policies, \texttt{load\_cached\_context} invalidated stale cache entries in all 8 probes (8/8).

\subsection{RQ4 \& RQ5: Impact Blast Radius, ATMS Slicing, Scaling, and Coordination Primitives}
\label{sec:exp_rq4_rq5}

\begin{figure}[t]
  \centering
  \includegraphics[width=\textwidth]{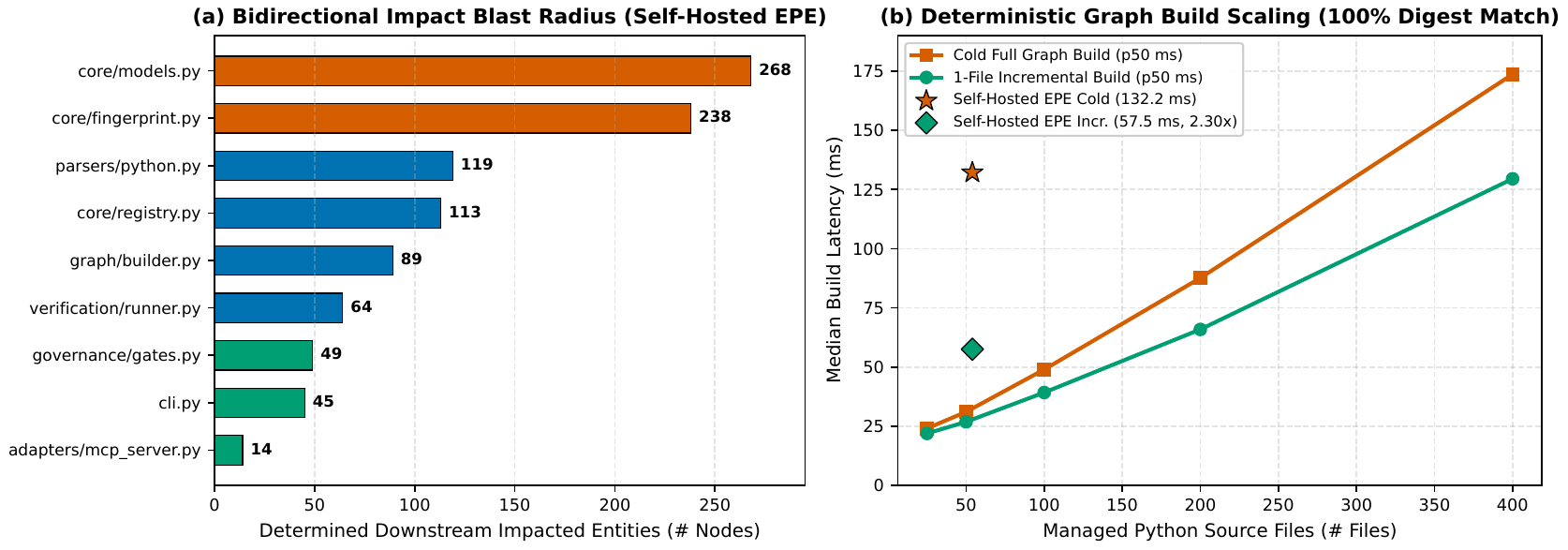}
  \caption{Impact propagation and graph scaling. (a) Determined downstream blast radius across self-hosted EPE source files from foundational \texttt{core} to leaf \texttt{adapters}. (b) Median latency scaling from 25 to 400 Python files alongside the denser 54-file self-hosted EPE repository, with exact SHA-256 digest equivalence preserved across all incremental updates.}
  \label{fig:scaling_impact}
\end{figure}

Figure~\ref{fig:scaling_impact}(a) quantifies why BTAE pairs bottom-up implementation with top-down contract governance. When running \texttt{analyze\_impact} on foundational \texttt{MOD-001} files such as \texttt{core/models.py} and \texttt{core/fingerprint.py}, reverse \texttt{depends\_on} and forward \texttt{defines}/\texttt{realizes} propagation deterministically reaches \textbf{268} and \textbf{238} downstream entities, respectively (8 module-level dependents), whereas leaf adapter \texttt{adapters/mcp\_server.py} affects only \textbf{14} entities and composition root \texttt{MOD-008} (\texttt{cli}) has 0 upstream module dependents. Consequently, BTAE's policy-derived risk floor concentrates mandatory human review where blast radius is largest rather than imposing uniform bureaucracy on leaf edits. Across 20 sequential single-file mutations on \texttt{src/epe/}, \texttt{build\_graph\_incremental} produced exact SHA-256 \texttt{content\_digest} matches with cold \texttt{build\_graph} in \textbf{20/20 (100\%)} trials, confirming Proposition~\ref{prop:incremental}. On synthetic multi-module Python repositories scaling from $N = 25$ to $N = 400$ files (117 to 1,617 nodes; Figure~\ref{fig:scaling_impact}(b), Table~\ref{tab:rq4_rq5_scaling} in Appendix~\ref{app:exp_tables}), racy-safe stat-cached incremental builds remain faster than cold builds across every scale, finishing at 129.53\,ms vs.\ 173.71\,ms at 400 files ($1.34\times$; on the denser 54-file self-hosted EPE repo, incremental reuse yields a \textbf{$2.30\times$} speedup from 132.18\,ms to 57.51\,ms). Furthermore, evaluating \texttt{compute\_assumption\_slice} across the 13 self-hosted module and class targets reduces the active ATMS evidence slice from all 189 verification-manifest files to a mean of \textbf{39.85 files (78.92\% reduction; 91.01\%--96.30\% on class targets)}, and \texttt{assess\_defeasible\_validity} correctly flags 10/10 upstream slice edits as \texttt{assumption\_undercut}, preserves \texttt{slice\_status = valid} across 10/10 disjoint leaf edits, and separates 10/10 \texttt{execution\_rebuttal} failures from 10/10 \texttt{assumption\_undercut} invalidations.

Table~\ref{tab:rq4_rq5_scaling} (right) in Appendix~\ref{app:exp_tables} also reports RQ5 overhead for the \texttt{events.jsonl} hash-chain audit log. Because each \texttt{append\_audit} acquires an exclusive \texttt{flock}, verifies the existing hash chain, writes canonical UTF-8 JSON, and calls \texttt{os.fsync}, per-event commit latency averages 2.44--4.87\,ms up to 500 events, while read-locked full-chain verification (\texttt{verify\_audit}) takes just \textbf{0.89\,ms} at 100 events and \textbf{4.45\,ms} at 500 events. In multi-agent coordination tests, exclusive POSIX ChangeSet leases (\texttt{acquire\_lease}, \texttt{renew\_lease}, \texttt{release\_lease}) deterministically blocked concurrent duplicate acquisitions via \texttt{LeaseConflict} and automatically recovered stale locks after TTL expiration (50\,ms probe). Similarly, \texttt{detect\_changeset\_conflicts} accurately flagged both \texttt{write\_scope}--\texttt{write\_scope} overlap (\texttt{src/epe/core/**} vs.\ \texttt{src/epe/core/models.py}) and \texttt{protected}--\texttt{write\_scope} overlap (\texttt{docs/architecture/**} vs.\ \texttt{docs/architecture/overview.md}) while leaving disjoint edits (\texttt{src/epe/cli.py}) unblocked.

\subsection{Limitations and Threats to Validity}
\label{sec:exp_limitations}

Consistent with BTAE's own anti-overclaiming charter, we explicitly delineate four boundaries of our current implementation and evaluation. First, EPE's static Python AST analyzer cannot resolve runtime reflection, dependency injection, or external RPC topology; our randomized measurements quantify this boundary rather than conceal it---22 of 40 dynamic-loading forms and all non-Python artifacts pass silently (Section~\ref{sec:exp_rq2}), while fail-closure conversely rejects all 15 probed standard-library lazy-loading tasks---so polyglot repositories and runtime-loaded modules require future Tree-sitter adapters and runtime trace integration. Second, while \texttt{derive\_risk\_floor} and \texttt{effective\_protected\_paths} eliminate local task-risk and ChangeSet-scope self-downgrades, \texttt{engineering/epe.yaml}, policy files, and EPE engine modules remain \emph{policy-boundary inputs}: as quantified in RQ2, an unisolated local process tampering with on-disk configuration or engine code degrades local protection from 11.07\% (CI-anchored) to 25.36\% FAR, and POSIX \texttt{flock}/\texttt{O\_NOFOLLOW} primitives assume POSIX semantics. Mandatory enforcement therefore requires pairing EPE with container/OS isolation, external secret-managed signing keys, and protected-branch CI gates (as configured on \texttt{lanshi17/BTAE}). Third, our empirical evaluation is self-hosted and protocol-focused: the self-hosted EKG contains 4 explicit \texttt{realizes} edges alongside 546 static \texttt{depends\_on} edges (testing declarative binding mechanics rather than dense repository-wide manual annotation), and we do not yet report agent-in-the-loop trajectories (e.g., SWE-bench multi-turn runs), third-party repository case studies, or longitudinal human-team productivity and defect-escape trials. Fourth, from a socio-technical and governance perspective, local-only execution provides a strong on-premise privacy advantage over cloud LLM judges, whereas mandatory human signing introduces key-gatekeeping bottlenecks for external contributors, review fatigue/rubber-stamping~\cite{bacchelli2013expectations}, and potential liability shifting onto individual signers; similarly, \texttt{C0}--\texttt{C3} context compaction does not automatically redact sensitive internal rules or commercial ADRs entering LLM prompts, and immutable \texttt{events.jsonl} audit logs warrant explicit organizational retention and access-control policies to prevent developer surveillance misuse.

\section{Related Work}
\label{sec:related_work}

\paragraph{Autonomous Coding Agents and Execution Scaffolds.}
Benchmarks for repository-level issue resolution such as SWE-bench~\cite{jimenez2024swebench} have catalyzed specialized Agent-Computer Interfaces (ACIs) including SWE-agent~\cite{yang2024sweagent}, sandboxed event-stream platforms such as OpenHands~\cite{wang2025openhands}, AST-guided fault localization in AutoCodeRover~\cite{zhang2024autocoderover}, and structured localization pipelines such as Agentless~\cite{xia2024agentless}. While these frameworks improve single-issue patch generation through shell tooling, file search, and unit-test feedback, they optimize per-task pass rates rather than multi-turn architectural continuity: none maintain persistent design-to-symbol realization graphs, enforce cryptographic approval bindings across edits, or distinguish whether an earlier passing test run has been invalidated by subsequent repository changes. BTAE complements agent execution scaffolds rather than replacing them, exposing read-only planning context via MCP~\cite{anthropic2024mcp} while enforcing scope baselines, capability-attested host adapters, and four-dimensional completion gates in EPE.

\paragraph{Engineering Knowledge Graphs and Traceability Recovery.}
Bridging high-level architectural intent and source implementation has been studied extensively in requirements traceability~\cite{gotel1994traceability,clelandhuang2014traceability}, software reflexion models~\cite{murphy1995reflexion,murphy2001reflexion}, and change impact analysis~\cite{bohner1996impact,lehnert2011impact}. Recent automated traceability approaches—such as transitive model-mediated linking in TransArC~\cite{keim2024transarc} and RAG-based link recovery in LiSSA~\cite{fuchss2025lissa}—infer documentation-to-code links via NLP and LLMs, while RepoDoc~\cite{xu2026repodoc} constructs repository knowledge graphs (\texttt{RepoKG}) to curb token inflation during automated documentation synthesis and incremental updates. However, probabilistic trace recovery and documentation generators treat source code as the primary truth to summarize and cannot serve directly as fail-closed authorization gates because heuristic or LLM-inferred edges carry non-zero false-positive risk. EPE instead records explicit origin types ($\{\texttt{explicit}, \texttt{static}, \texttt{inferred}\}$), publishes lock-protected SHA-256 graph generations atomically, labels uncertain impact paths as \texttt{candidate}, and forbids unverified inferences from satisfying design or completion gates; its ATMS assumption-slicing and defeasible validity engine directly instantiates Assumption-Based Truth Maintenance Systems (ATMS)~\cite{dekleer1986atms} and Pollock's distinction between \emph{undercutting} and \emph{rebutting} defeaters~\cite{pollock1987defeasible} over working-tree byte assumptions.

\paragraph{Specification-Driven Development and Contract Grounding.}
Design by Contract~\cite{meyer1992dbc} and consumer-driven contract testing frameworks such as Pact~\cite{pact2024docs} formalize behavioral obligations across component interfaces, and recent empirical work at Google demonstrates that grounding AI agents in extracted pre-/post-condition contracts improves bug-detection rates in automated test generation~\cite{tufano2025contracts}. In parallel, specification-driven agent toolkits such as GitHub Spec Kit~\cite{github2025speckit} and AWS Kiro~\cite{aws2025kiro} introduce repository constitutions, steering files, and upfront specification templates. Despite these advances, existing specification toolkits typically assume uniform feature-first workflows and rely on mutable Markdown checkboxes or frontmatter flags that an agent can edit directly without cryptographic verification. BTAE supports four distinct workflow entrypoints (\texttt{feature}, \texttt{bugfix}, \texttt{refactor}, and \texttt{exploration}) under machine-validated JSON Schemas and binds design approvals to SHA-256 digests encompassing both the design prose and all referenced Draft 2020-12 contract schemas.

\paragraph{Architectural Fitness Functions, Policy-as-Code, and Evidentiary Provenance.}
Evolutionary architecture governance advocates executable architectural fitness functions inside deployment pipelines~\cite{ford2022evolutionary} paired with Architecture Decision Records (ADRs)~\cite{nygard2011adr}, while policy-as-code engines such as Open Policy Agent (OPA)~\cite{opa2024docs} decouple policy evaluation from application services and software supply-chain frameworks such as in-toto~\cite{torresarias2019intoto} and Sigstore~\cite{newman2022sigstore} bind build steps and artifacts to cryptographic signatures and transparency logs. Conventional fitness linters, however, treat architectural exceptions as static ignore lists that silently outlive their intended temporary window and treat dynamic-import blind spots as absence of violation, while empirical studies of modern code review show that human approval can become perfunctory under review fatigue~\cite{bacchelli2013expectations}. EPE unifies architectural fitness with local evidentiary provenance: ADR waivers require \texttt{approved} entity status, exact $(\texttt{rule\_id}, \texttt{from}, \texttt{to})$ scope matching, and an unexpired UTC \texttt{review\_until} deadline; AST blind spots fail closed when \texttt{unknown\_is\_blocking} is enabled; subprocess JUnit evidence is bound to pre-/post-run SHA-256 input manifests and an append-only hash-chain audit log; and cryptographic human approval is reserved for medium- and high-risk changes to bound review overhead while establishing verifiable signer provenance.

\paragraph{Content-Addressed Build Systems and Git-Native Alternatives.}
Content-addressed deployment and build systems such as Nix~\cite{dolstra2004nix} and Reproducible Builds~\cite{lamb2022reproducible} establish bit-for-bit integrity by indexing derivations by input cryptographic hashes. A natural question is why BTAE maintains its own SHA-256 manifests and generation snapshots rather than relying solely on \texttt{git write-tree} Merkle hashes and signed Git tags. Three protocol requirements motivate EPE's entity-level layer: (i) Git content-addresses committed or staged whole files, whereas EPE tracks logical entities and AST-qualified symbols across file splits and renames; (ii) coding agents iterate over dirty, uncommitted working trees where pre-/post-test manifests must exclude derived governance indices (\texttt{.engineering/}, \texttt{reports/}) without polluting Git commit history; and (iii) EPE differentiates edge provenance (\texttt{explicit}/\texttt{static}/\texttt{inferred}) and binds multi-schema contract revisions ($H_{\text{rev}}(e)$) and JUnit execution counts inside a zero-daemon local engine. In Section~\ref{sec:exp_rq2}, we empirically compare EPE against both a minimal Git tree-hash/signed-tag script (\textbf{B\textsubscript{git}}) and a combined Git+CI+static+CODEOWNERS stack (\textbf{B3}) to quantify these incremental protocol gains.

\section{Conclusion}
\label{sec:conclusion}

This paper addressed long-horizon design drift, silent architectural erosion, and completion-gate spoofing in agentic software maintenance by introducing Bottom-Up Traceable Agentic Engineering (BTAE) and its local deterministic reference engine, the Engineering Protocol Engine (EPE). By separating provenance-typed facts in an atomic Directory-Merkle Engineering Knowledge Graph (EKG) from authorization in a fail-closed Engineering Governance Plane (EGP) operating over an orthogonal four-dimensional epistemic state $\mathcal{S}(e) = (d, m, r, v)$, EPE combines policy-derived blast-radius risk floors with ATMS-sliced defeasible validity so that---under anchored configuration and trust-root boundaries---self-reported frontmatter, downgraded task risk declarations, static existence checks, and stale test runs cannot masquerade as valid completion evidence. Empirical evaluation on the self-hosted 7,338-LOC EPE codebase (426 EKG nodes, 550 edges, 368 passing tests) and a 620-trial randomized fault-injection suite demonstrated 10/10 byte-identical graph reconstructions at 132.2\,ms cold and 57.5\,ms incremental median latency ($2.30\times$ median speedup via 50\,ms racy-safe stat caching), reduced conditional false acceptance from 49.46\% in a strong Git+CI+static stack (\textbf{B3}) and 83.75--100.00\% in narrower baselines to \textbf{11.07\%} (95\% CI 8.73--13.94) under anchored CI policy boundaries (25.36\% with local risk floors, 32.50\% unanchored) with 0/60 false rejections on standard compliant workflows, compressed task context payloads by \textbf{97.57\%} on average relative to full-repository loading (8/8 single-byte cache invalidation probes), and narrowed evidence invalidation slices by \textbf{78.92\%} while separating \texttt{undercut} from \texttt{rebutted} failures. The measured misses---reflective dynamic-import evasion forms unknown to the static analyzer and non-Python artifacts outside AST coverage, together with the fail-closed rejection of all 15 probed standard-library lazy-loading tasks---delineate the protocol's current boundary and define the runtime-trace and polyglot-parser extensions that future work must close.

The primary scope boundaries of the present system lie in its static Python AST fact extractor, its reliance on CI/OS boundaries to protect \texttt{engineering/epe.yaml}, policy files, and engine modules against unilateral local tampering, and its self-hosted evaluation scope. Future work will extend EPE along four directions: (i) multi-language AST extraction via Tree-sitter paired with runtime OpenTelemetry trace reconciliation to narrow dynamic blind spots; (ii) container-enforced host sandboxes and remote append-only transparency logs implementing \texttt{HostPort v1} write isolation alongside signed task/scope delegation; (iii) multi-turn game-theoretic evaluation of governance-arbitrage (``salami-slicing'') defenses under derived risk floors; and (iv) agent-in-the-loop evaluations (e.g., SWE-bench and third-party open-source repositories) plus controlled longitudinal trials measuring defect escape rates, review effort, and token economics across multi-developer teams.

\section*{Acknowledgments}
This work was supported in part by the Hunan Provincial Natural Science Foundation under Grant 2026JJ80034, in part by the New Engineering Research and Practice Project of the Hunan Provincial Department of Education under Grant 202503000046, and in part by the China Scholarship Council under Grant CXXM2510173682. The complete reference implementation, JSON Schemas, self-hosted governance artifacts, and reproducible evaluation scripts are available in the companion artifact repository.

\bibliographystyle{IEEEtran}
\bibliography{references}

\clearpage
\appendix
\section{Proofs of Formal Properties}
\label{app:proof}

\begin{proof}[Proof of Proposition~\ref{prop:incremental}]
Let $\mathcal{W}'$ be the current repository state under parser configuration $\pi = (\texttt{version},\allowbreak \texttt{builder\_version},\allowbreak \texttt{python},\allowbreak \texttt{code\_roots},\allowbreak \texttt{module\_roots})$, and let $G_{\text{prev}}$ be a previously published snapshot. Before reusing any cached data, \texttt{\_validated\_previous} checks four preconditions: (1) $G_{\text{prev}}.\texttt{complete} = \texttt{true}$ and $G_{\text{prev}}.\texttt{parser\_config} = \pi$; (2) $G_{\text{prev}}.\texttt{content\_digest}$ equals the canonical SHA-256 digest of $G_{\text{prev}}$ excluding \texttt{content\_digest}; (3) $G_{\text{prev}}.\texttt{input\_manifest}$ contains strictly sorted unique pairs $(p, h_{\text{prev}}(p))$ matching its stored manifest digest; and (4) every stored \texttt{CodeParseResult} $f \in G_{\text{prev}}.\texttt{code\_facts}$ round-trips identically with $f.\texttt{complete} = \texttt{true}$ and $f.\texttt{digest} = h_{\text{prev}}(f.\texttt{path})$ for all managed Python paths in $\mathcal{M}_{\text{prev}}$. If any precondition fails, \texttt{build\_graph\_incremental} executes a full \texttt{parse\_python\_tree} and delegates directly to \texttt{build\_graph}, making equivalence immediate.

Otherwise, let $\mathcal{M}_{\text{curr}}$ be the sorted SHA-256 manifest of $\mathcal{W}'$ (which fails closed if any file mutates during build via the pre-/post-manifest check $\mathcal{M}_{\text{curr}} = \mathcal{M}'_{\text{curr}}$), and let $\Delta = \{p \in \mathcal{M}_{\text{prev}} \cup \mathcal{M}_{\text{curr}} \mid h_{\text{prev}}(p) \neq h_{\text{curr}}(p)\}$. Because \texttt{parse\_python(bytes, path)} is a pure deterministic function of file bytes, relative path, and CPython major/minor AST grammar (captured in $\pi$), every unchanged managed Python file $p \notin \Delta$ satisfies $\text{bytes}_{\text{curr}}(p) = \text{bytes}_{\text{prev}}(p)$ (up to SHA-256 collision resistance) and therefore yields an identical \texttt{CodeParseResult} $f_p$. Deleted files ($p \in \Delta \setminus \mathcal{M}_{\text{curr}}$) are excluded from the reused set, and added or modified Python files ($p \in \Delta \cap \mathcal{M}_{\text{curr}}$) are parsed fresh via \texttt{parse\_python}. In \texttt{build\_graph}, \texttt{code\_results} is sorted canonically by \texttt{item.path} before import resolution, module aggregation, SHA-256 edge deduplication, and lexicographic node/edge sorting; furthermore, no wall-clock timestamp enters the snapshot dictionary. Hence the union $\mathcal{F}_{\text{reused}} \cup \mathcal{F}_{\text{reparsed}}$ after sorting by \texttt{path} is byte-for-byte identical to the list produced by cold \texttt{parse\_python\_tree}$(\mathcal{W}')$, yielding identical canonical JSON payloads and identical SHA-256 \texttt{content\_digest} values.
\end{proof}

\begin{proof}[Proof of Lemma~\ref{lem:cache_freshness}]
By construction of \texttt{store\_cached\_context}, any bundle with $\texttt{incomplete} = \texttt{true}$---whether triggered by budget omissions, per-item character truncation, or mid-assembly workspace mutation---is rejected prior to serialization and never written to the cache directory. Entries are indexed by \texttt{context\_cache\_key} (v6), which hashes the selection-policy version, budget $B$, serialized task $T$, required-source list and byte digests, source assessments, snapshot \texttt{content\_digest}, and full \texttt{context\_inputs\_manifest} digest (covering managed design, code, test, and contract roots plus \texttt{AGENTS.md}, \texttt{archive/}, \texttt{docs/architecture/decisions/}, \texttt{engineering/epe.yaml}, policies, approvals, audit logs, and evidence). On lookup, \texttt{load\_cached\_context} recomputes both \texttt{context\_cache\_key} and \texttt{context\_inputs\_manifest} over $\mathcal{W}$, verifies the stored \texttt{content\_digest}, and re-hashes every cached item file against its stored \texttt{content\_sha256}. Under SHA-256 collision resistance, a cache hit guarantees $\texttt{incomplete} = \texttt{false}$, exact selection-policy equality, and byte-for-byte freshness of every cached item.
\end{proof}

\begin{proof}[Proof of Lemma~\ref{lem:impact_termination}]
Each node $u \in \mathcal{V}$ reaches one of at most three ordered label states: $\texttt{unvisited} \prec \texttt{candidate} \prec \texttt{determined}$. Traversal inserts a node into the frontier only upon a strict upward transition (either first discovery or promotion from \texttt{candidate} to \texttt{determined}) and stops admitting new distinct nodes once the visited set reaches $B_{\text{imp}}$. Because labels increase strictly along a finite lattice of height 2, each visited node can trigger outgoing edge propagation at most twice (once when discovered as \texttt{candidate}, and at most once more if later promoted to \texttt{determined}), guaranteeing termination in $O(|\mathcal{V}| + |\mathcal{E}|)$ edge inspections while bounding visited nodes by $\min(B_{\text{imp}}, |\mathcal{V}|)$.
\end{proof}

\begin{proof}[Proof of Lemma~\ref{lem:validity_monotonicity}]
During \texttt{evidence\_validity}, EPE recomputes from disk the SHA-256 digests of every artifact in $E^*$, the verification policy configuration $h'_{\text{pol}}$, runtime environment metadata and \texttt{max\_age\_seconds}, and the current verification manifest $\mathcal{M}_{\text{ver}}(\mathcal{W}')$. Under SHA-256 collision resistance: (1) any byte mutation or missing file among $E^*$'s recorded artifacts yields an artifact digest mismatch and classifies $E^*$ as \texttt{invalid}; (2) missing/corrupted provenance or runtime/clock skew classifies $E^*$ as \texttt{unknown}; and (3) any single-byte modification, addition, or deletion in any path $p \in \text{dom}(\mathcal{M}_{\text{ver}}(\mathcal{W})) \cup \{\text{policy}(\mathcal{C})\}$ yields $H(\mathcal{M}_{\text{ver}}(\mathcal{W}')) \neq H(\mathcal{M}_{\text{ver}}(\mathcal{W}))$ or $h'_{\text{pol}} \neq h_{\text{pol}}$, classifying $E^*$ as \texttt{stale}. In every case, $v(E^*, \mathcal{W}') \neq \texttt{valid}$. Moreover, when \texttt{compute\_assumption\_slice} succeeds without AST blind spots or missing targets, \texttt{assess\_defeasible\_validity} checks $\Delta_{\text{slice}} = \{p \in \mathcal{A}(E_0, G) \mid h_{\text{old}}(p) \neq h_{\text{curr}}(p)\}$; any mutation inside $\mathcal{A}(E_0, G)$ yields $\Delta_{\text{slice}} \neq \emptyset$ and hence $v_{\text{def}} = \texttt{undercut}$.
\end{proof}

\begin{proof}[Proof of Lemma~\ref{lem:gate_soundness}]
Under assumptions (i)--(iii), we verify each in-coverage family against Algorithm~\ref{alg:finalize}:
\begin{enumerate}[leftmargin=1.6em,itemsep=1pt,topsep=2pt]
  \item \emph{Approval forgery / staleness} ($F_1$--$F_2$): Line~3 validates the \texttt{events.jsonl} SHA-256 chain; Line~4 verifies OpenSSH Ed25519 signatures over $H_{\text{rev}}(e)$ under trusted keys, namespace \texttt{epe-approval}, and required risk/human constraints. Any unsigned, wrong-namespace, untrusted, tampered, or post-edit document fails \texttt{design\_ok}.
  \item \emph{Mapping drift} ($F_3$): \texttt{load\_registry} rejects duplicate IDs, duplicate YAML keys, and dangling relations; Line~6 (\texttt{ValidateASTMapping}) detects missing/renamed AST symbols and fails \texttt{mapping\_ok}.
  \item \emph{ADR lifecycle / static forbidden / direct dynamic imports} ($F_4, F_9, F_{10}$): Line~7 validates \texttt{ADR-*} status (\texttt{approved}), future UTC \texttt{review\_until}, and $(\texttt{rule\_id}, \texttt{from}, \texttt{to})$ matching, failing \texttt{arch\_ok} on forbidden edges and blocking \texttt{EPE-AST-DYNAMIC} under \texttt{unknown\_is\_blocking = true}.
  \item \emph{Evidence faults} ($F_5$): Lines~9--15 reject empty check policies, require timezone-aware UTC ordering, recompute artifact/policy/manifest SHA-256s and JUnit counts ($\ge 1$ passed, $0$ failed) via Lemma~\ref{lem:validity_monotonicity}, and enforce $v(E^*) = \texttt{valid}$.
  \item \emph{Scope escape / audit tamper / host attestation} ($F_6$--$F_8$): Line~8 checks \texttt{CompareBaseline} against $\texttt{write\_scope} \setminus \texttt{effective\_protected\_paths}$; \texttt{verify\_audit} detects chain breaks or reorderings; and \texttt{HermesHostAdapter} rejects unsigned, replayed-nonce, escalated-op, or foreign-key capabilities.
  \item \emph{Risk downgrade / local policy or engine tampering} ($F_{11}$--$F_{12}$): Line~2 computes $r_{\text{eff}}(T) = \max(T.\texttt{risk}, \rho_{\text{floor}})$, blocking downgraded \texttt{T.risk} via \texttt{EPE-RISK-DOWNGRADE} and escalating Line~4 to $r_{\text{eff}}(T)$; assumption (i) prevents local tampering with \texttt{epe.yaml}, policies, or engine code.
\end{enumerate}
Hence every instance in $F_1$--$F_{12}$ triggers at least one blocking/error diagnostic in $\mathcal{D}$.
\end{proof}

\section{Acceptance Traceability Matrix and CLI Exit Contract}
\label{app:ac_matrix}

Table~\ref{tab:ac_matrix} summarizes the twelve specification acceptance criteria (\texttt{AC-01}--\texttt{AC-12}) enforced in EPE's automated regression suite, mapping each requirement to its failure-mode assertion and verification test entrypoint. All CLI commands emit schema-validated \texttt{CommandResult} JSON objects on \texttt{stdout} (keeping execution logs on \texttt{stderr}) and map outcomes to deterministic POSIX exit codes: \texttt{0} (\texttt{passed}), \texttt{1} (\texttt{blocked} by governance gate or safe-init conflict), \texttt{2} (parameter, policy, or schema configuration error), \texttt{3} (execution, parser, or storage fault), \texttt{4} (\texttt{unsupported} operation or host capability), and \texttt{5} (concurrent \texttt{EPE-WRITE-CONFLICT}).

\begin{table}[!htbp]
  \centering
  \caption{Traceability matrix of specification acceptance criteria (\texttt{AC-01}--\texttt{AC-12}) in the self-hosted EPE suite.}
  \label{tab:ac_matrix}
  \footnotesize
  \begin{tabularx}{\textwidth}{llX}
    \toprule
    ID & Subsystem & Verified Positive \& Fail-Closed Negative Assertions \\
    \midrule
    \texttt{AC-01} & Safe Init & Creates missing governance files; pre-checks conflicts and symlinks without overwriting existing bytes. \\
    \texttt{AC-02} & Registry & Rejects duplicate entity IDs, duplicate YAML keys, unknown fields, and dangling or self-looping relations. \\
    \texttt{AC-03} & AST Mapping & Validates multi-file/multi-symbol \texttt{realizes} links; blocks when an \texttt{implemented} symbol is deleted. \\
    \texttt{AC-04} & Architecture & Detects module/source dependency cycles and forbidden globs; surfaces \texttt{EPE-AST-DYNAMIC} blind spots. \\
    \texttt{AC-05} & Design Gate & Blocks missing designs, stale approval revisions, unauthorized high-risk changes, and out-of-scope edits. \\
    \texttt{AC-06} & Test Runner & Distinguishes passed, failed, zero-test JUnit collection, process-group timeout, and missing commands. \\
    \texttt{AC-07} & Provenance & Captures dirty-tree/commit-null SHA-256 manifests; downgrades evidence to \texttt{stale} after 1-byte edits. \\
    \texttt{AC-08} & Validity & Marks corrupted artifacts \texttt{invalid} and missing/mismatched Python or OS metadata \texttt{unknown}. \\
    \texttt{AC-09} & CLI Contract & Enforces \texttt{command-result.schema.json}, exit codes \texttt{0}--\texttt{5}, and clean \texttt{stdout}/\texttt{stderr} separation. \\
    \texttt{AC-10} & Atomic EKG & Verifies cold-rebuild SHA-256 idempotence, atomic \texttt{generations/} rename, and lock conflict code \texttt{5}. \\
    \texttt{AC-11} & Workflows & Validates \texttt{feature}, \texttt{bugfix}, \texttt{refactor}, and \texttt{exploration} mandatory fields before task activation. \\
    \texttt{AC-12} & Closed Loop & Executes end-to-end design $\rightarrow$ code $\rightarrow$ verify $\rightarrow$ check loop; blocks mid-run input mutations. \\
    \bottomrule
  \end{tabularx}
\end{table}
\FloatBarrier

\section{Randomized Fault Family Catalog and Measured Outcomes}
\label{app:scenarios}

Table~\ref{tab:scenarios_summary} catalogs the fault families instantiated by the seeded randomized suite of Section~\ref{sec:exp_rq2}: 14 adversarial families (40 trials each, $N = 560$) plus the valid-workflow class (60 trials). All fault parameters are drawn from fixed seed \texttt{20260520}, and columns report measured correct-decision counts per governance regime/mode. Both out-of-coverage boundary families are drawn deliberately outside the static Python AST protocol to measure blind spots rather than conceal them.

\begin{table}[!htbp]
  \centering
  \caption{Fault family catalog and measured correct-decision counts per governance regime in RQ2 (40 trials per fault family, 60 valid workflows). \textbf{EPE\textsubscript{flr}}: local policy-derived risk floor; \textbf{EPE\textsubscript{full}}: CI/policy-anchored EPE.}
  \label{tab:scenarios_summary}
  \footnotesize
  \setlength{\tabcolsep}{3.5pt}
  \begin{tabularx}{\textwidth}{lXrrrrrrr}
    \toprule
    Family ($n$) & Representative randomized variants & B0 & B1 & B2 & B\textsubscript{git} & B3 & EPE\textsubscript{flr} & EPE\textsubscript{full} \\
    \midrule
    approval-forgery (40) & self-declared, wrong namespace/payload, tampered sig, untrusted signer, narrow risk & 0 & 0 & 0 & 0 & 0 & 40 & 40 \\
    approval-staleness (40) & design doc mutated after signing (prose, entity, rollback) & 0 & 0 & 0 & 40 & 40 & 40 & 40 \\
    mapping-drift (40) & deleted/renamed symbol, dup ID/YAML key, dangling relation & 0 & 40 & 0 & 0 & 40 & 40 & 40 \\
    adr-lifecycle (40) & draft, expired, and scope-mismatched ADR waivers & 0 & 11 & 0 & 0 & 11 & 40 & 40 \\
    evidence-fault (40) & post-test edit, artifact tamper, env/age/zero-test, failed run & 0 & 0 & 9 & 16 & 16 & 40 & 40 \\
    scope-escape (40) & writes outside baseline and into protected paths & 0 & 0 & 0 & 21 & 40 & 40 & 40 \\
    audit-tamper (40) & log byte flip, raw append breaking hash chain, reorder & 0 & 0 & 0 & 0 & 0 & 40 & 40 \\
    host-attestation (40) & unsigned capability, replayed nonce, escalated ops, foreign key & 0 & 0 & 0 & 0 & 0 & 40 & 40 \\
    static-forbidden (40) & forbidden cross-module import in randomized core file & 0 & 40 & 0 & 0 & 40 & 40 & 40 \\
    risk-downband (40) & contract/arch/blast/policy-protected edit declared \texttt{risk=low} & 0 & 0 & 0 & 0 & 16 & 40 & 40 \\
    config-tamper (40) & local disable of risk floor/\texttt{unknown\_is\_blocking}, rogue trust key & 0 & 0 & 0 & 0 & 40 & 0 & 40 \\
    engine-tamper (40) & local stubbing of gate/verifier/approval/scope checks & 0 & 0 & 0 & 0 & 40 & 0 & 40 \\
    \midrule
    dynamic: direct (18) & \texttt{importlib.import\_module}, \texttt{\_\_import\_\_}, concat string & 0 & 0 & 0 & 0 & 0 & 18 & 18 \\
    dynamic: reflective (22) & \texttt{find\_spec}, aliased \texttt{import\_module}, \texttt{sys.modules} & 0 & 0 & 0 & 0 & 0 & 0 & 0 \\
    nonpython-artifact (40) & \texttt{.js}/\texttt{.json}/\texttt{.md} referencing forbidden modules & 0 & 0 & 0 & 0 & 0 & 0 & 0 \\
    \midrule
    valid-workflow (60) & plain, risk-randomized, and validly ADR-waived workflows & 60 & 60 & 60 & 60 & 60 & 60 & 60 \\
    \bottomrule
  \end{tabularx}
\end{table}
\FloatBarrier

\section{Host Capability Matrix and Agent Integration Surfaces}
\label{app:host_matrix}

To prevent coding agents or orchestrators from mistaking transport connectivity for security isolation, EPE explicitly separates read-only planning surfaces from execution-host capabilities:
\begin{enumerate}[leftmargin=1.8em,itemsep=2pt,topsep=3pt]
  \item \textbf{Read-Only Stdio MCP Entrypoint (\texttt{btae-mcp}):} Implements JSON-RPC 2.0 over \texttt{stdin}/\texttt{stdout} supporting MCP protocol versions \texttt{2024-11-05}, \texttt{2025-03-26}, and \texttt{2025-06-18}. Exposes three read-only tools (\texttt{btae\_project\_context}, \texttt{btae\_design\_plan}, \texttt{btae\_validate\_project}; all annotated \texttt{readOnlyHint:\ true}) and eight authoritative resources (\texttt{btae://charter}, \texttt{btae://spec}, \texttt{btae://architecture/*}, \texttt{btae://workflow/*}), capping per-resource reads at 32\,KB and rejecting symlink traversal.
  \item \textbf{HostPort v1 Capability States:} \texttt{local-cli} exposes process-group timeout control and read-only \texttt{verify} dispatch (\texttt{status:\ measured}), returning \texttt{unsupported} for \texttt{start\_implementation} and \texttt{finalize}. \texttt{HermesACPAdapter} provides bounded POSIX \texttt{initialize}, \texttt{session/new} (forcing \texttt{mcpServers:\ []}), single-reader \texttt{session/update} streaming, and registered child-session forking (\texttt{fork\_session}), while remaining \texttt{handshaken-untrusted} with \texttt{trusted\_identity:\ false} and \texttt{write\_isolation:\ false}. \texttt{HermesHostAdapter} transitions from \texttt{handshaken-untrusted} to \texttt{handshaken} only when the host signs the canonical JSON challenge payload containing \texttt{protocol\_version:\ "1"}, \texttt{nonce}, and \texttt{capabilities} using an OpenSSH Ed25519 key listed in \texttt{engineering/epe.yaml} (\texttt{namespace="epe-host-capability"}) and requests a subset of its configured \texttt{operations}.
\end{enumerate}

\section{Self-Hosted Repository and EKG Composition}
\label{app:repo}

The reference implementation evaluated throughout this paper is the self-hosted EPE repository (commit \texttt{d67f9295}): 55 Python source files (7,338 lines of code: 54 managed files across 9 logical modules \texttt{MOD-001}--\texttt{MOD-009} under \texttt{src/epe/} plus the stdio \texttt{btae-mcp} entrypoint), 15 Draft 2020-12 JSON Schemas, 4 workflow definitions, 17 registered architectural/class/contract entities, and 74 test files (27 unit, 21 integration, and 26 acceptance suites covering \texttt{AC-01}--\texttt{AC-12}). Its published Engineering Knowledge Graph (\texttt{graph.json}: 519.67\,KB; \texttt{entities.json}: 9.03\,KB) contains \textbf{426 nodes} (355 AST \texttt{Symbol} nodes, 54 \texttt{Source} files, 9 \texttt{Module} entities, 4 core \texttt{Class} entities, 1 \texttt{Architecture}, 1 \texttt{Contract}, 1 \texttt{Intent}, and 1 \texttt{Spec}) and \textbf{550 provenance-typed edges} (4 explicit \texttt{realizes} mappings and 546 static \texttt{depends\_on} edges comprising 164 internal file imports, 116 module-aggregated dependencies, and 266 external package references).

\section{Detailed Operation Latencies}
\label{app:latency}

Table~\ref{tab:rq1_latency} reports per-operation execution latencies across $N = 25$ independent trials on the self-hosted repository, separating advisory interactive operations from governance gate/batch execution. Cache-hit latency is measured on a complete non-truncated bundle.
\begin{table}[t]
  \centering
  \caption{Execution latencies ($N = 25$ trials) on the self-hosted 7,338-LOC repository (426 nodes, 550 edges), separating advisory interactive checks from governance gate/batch runs.}
  \label{tab:rq1_latency}
  \resizebox{\textwidth}{!}{%
  \begin{tabular}{llrrrr}
    \toprule
    Operation & Primary Subsystem & p50 (ms) $\downarrow$ & Mean $\pm$ SD (ms) $\downarrow$ & p95 (ms) $\downarrow$ & Min / Max (ms) \\
    \midrule
    \multicolumn{6}{l}{\textit{Advisory Interactive Operations}} \\
    \texttt{policy validate} & Schema \& Workflow Policy & 2.04 & $2.05 \pm 0.06$ & 2.10 & 1.99 / 2.27 \\
    \texttt{registry load} & Markdown/YAML Frontmatter & 8.10 & $8.19 \pm 0.38$ & 9.16 & 7.93 / 9.45 \\
    \texttt{mapping validate} & AST Symbol Realization & 74.57 & $74.84 \pm 2.20$ & 80.00 & 71.53 / 81.10 \\
    \texttt{graph build} (cold) & Full AST + Atomic Publish & 132.18 & $132.52 \pm 3.63$ & 137.87 & 127.09 / 138.90 \\
    \texttt{graph build} (1-file incr.) & Racy-Safe Stat + AST Reuse & \textbf{57.51} & $61.35 \pm 21.30$ & 61.96 & 53.49 / 162.98 \\
    \texttt{architecture validate} & Cycles + Import Boundaries & 75.46 & $75.18 \pm 3.36$ & 80.45 & 71.35 / 86.63 \\
    \texttt{impact analyze} & Bidirectional Traversal & 4.28 & $4.41 \pm 0.47$ & 4.83 & 4.14 / 6.47 \\
    \texttt{context resolve} (cold) & Stratified \texttt{C0}--\texttt{C3} Assembly & 15.37 & $15.54 \pm 0.66$ & 16.00 & 15.11 / 18.55 \\
    \texttt{context resolve} (cached) & SHA-256 Manifest Cache Hit & \textbf{1.98} & $1.99 \pm 0.04$ & 2.06 & 1.94 / 2.15 \\
    \midrule
    \multicolumn{6}{l}{\textit{Governance Gate \& Verification Batch Operations}} \\
    \texttt{epe check} (gate-only) & Composite G1--G5 + ATMS Check & 57.83 & $57.40 \pm 23.58$ & 93.34 & 16.90 / 94.34 \\
    \texttt{verify + check} (mini suite) & Subprocess \texttt{pytest} + Gate & 548.75 & $548.79 \pm 22.26$ & 575.08 & 512.69 / 596.41 \\
    \bottomrule
  \end{tabular}%
  }
\end{table}

\section{Context Assembly, Scaling, and Audit Performance Detail}
\label{app:exp_tables}

Tables~\ref{tab:rq3_context} and~\ref{tab:rq4_rq5_scaling} give the full numerical detail behind RQ3--RQ5: bounded context assembly across budgets and maintenance tasks, synthetic graph construction scaling from 25 to 400 files under racy-safe stat caching, and hash-chain audit log performance from 10 to 500 committed events.
\begin{table}[t]
  \centering
  \caption{Bounded context assembly efficiency across 13 representative tasks (9 module tasks and 4 class tasks) versus full-repository loading (190 files, 1,964.3\,KB, $\approx 491.1\text{k}$ tokens).}
  \label{tab:rq3_context}
  \resizebox{\textwidth}{!}{%
  \begin{tabular}{lrrrrrrr}
    \toprule
    Context Regime & Mean Items & Payload (KB) $\downarrow$ & Est. Tokens $\downarrow$ & Reduction (\%) $\uparrow$ & \texttt{C0} Ret. (13/13) $\uparrow$ & Recall (13/13) $\uparrow$ & Mean Omissions \\
    \midrule
    Full Repository & 190.00 & 1964.29 & 491,073 & 0.00 & 13/13 & 13/13 & 0.00 \\
    EPE Bounded ($B = 8$) & 7.00 & 40.81 & 10,203 & \textbf{97.92} & \textbf{13/13} & \textbf{13/13} & 6.23 \\
    EPE Bounded ($B = 16$) & 8.15 & 47.75 & 11,937 & 97.57 & \textbf{13/13} & \textbf{13/13} & \textbf{0.00} \\
    EPE Bounded ($B = 32$) & 8.15 & 47.75 & 11,937 & 97.57 & \textbf{13/13} & \textbf{13/13} & \textbf{0.00} \\
    \cmidrule(lr){1-8}
    \quad Module Tasks (9, $B=32$) & 9.44 & 58.92 & 15,085 & 96.93 & 13/13 & 13/13 & 0.00 \\
    \quad Class Tasks (4, $B=32$) & 5.25 & 18.96 & 4,854 & \textbf{99.01} & 13/13 & 13/13 & 0.00 \\
    \bottomrule
  \end{tabular}%
  }
\end{table}
\begin{table}[t]
  \centering
  \caption{Synthetic graph build scaling ($N = 25$--$400$ Python files) and append-only SHA-256 hash-chain audit log performance ($L = 10$--$500$ committed events with \texttt{flock} + \texttt{fsync}).}
  \label{tab:rq4_rq5_scaling}
  \resizebox{\textwidth}{!}{%
  \begin{tabular}{rrrrr@{\hspace{2.0em}}rrrr}
    \toprule
    \multicolumn{5}{c}{EKG Construction Scaling (RQ4)} & \multicolumn{4}{c}{Hash-Chain Audit Log Scaling (RQ5)} \\
    \cmidrule(lr){1-5} \cmidrule(lr){6-9}
    Files & Nodes / Edges & Cold p50 (ms) $\downarrow$ & Incr. p50 (ms) $\downarrow$ & Speedup $\uparrow$ & Events $L$ & Log Size (KB) & Append/Ev (ms) $\downarrow$ & Verify p50 (ms) $\downarrow$ \\
    \midrule
    25 & 117 / 42 & 24.05 & 21.93 & $1.10\times$ & 10 & 3.35 & 2.84 & 0.10 \\
    50 & 217 / 86 & 31.03 & 26.83 & $1.16\times$ & 50 & 16.83 & 2.44 & 0.44 \\
    100 & 417 / 174 & 48.94 & 39.21 & $1.25\times$ & 100 & 33.68 & 3.78 & 0.89 \\
    200 & 817 / 350 & 87.60 & 65.88 & $1.33\times$ & 500 & 169.28 & 4.87 & 4.45 \\
    400 & 1,617 / 700 & 173.71 & 129.53 & $1.34\times$ & --- & --- & --- & --- \\
    \bottomrule
  \end{tabular}%
  }
\end{table}

\section{Composite Finalize Gate Pseudocode}
\label{app:gate_algo}

Algorithm~\ref{alg:finalize} details the fail-closed verification-validity and composite finalize gate, whose five orthogonal invariants (G1--G5) are summarized in Section~\ref{sec:method_governance}.
\begin{algorithm}[t]
\caption{Fail-Closed Verification Validity and Composite Finalize Gate (\texttt{epe check})}
\label{alg:finalize}
\begin{algorithmic}[1]
\REQUIRE Repository root $\mathcal{W}$, config $\mathcal{C}$, task $T$, registry $\mathcal{E}$, fresh EKG snapshot $G$
\ENSURE Gate status $\in \{\texttt{passed}, \texttt{blocked}, \texttt{error}\}$, diagnostics $\mathcal{D}$
\STATE $\texttt{design\_ok} \leftarrow \bigwedge_{e \in T.\texttt{targets}} \text{ValidateDesign}(\mathcal{E}, e, \mathcal{W})$ \COMMENT{Includes local-only JSON Schema \texttt{\$ref} check}
\STATE $r_{\text{eff}} \leftarrow \max(T.\texttt{risk}, \rho_{\text{floor}}(T, \Delta\mathcal{W}, G))$ if $\mathcal{C}.\texttt{enforce\_risk\_floor}$ else $T.\texttt{risk}$ \COMMENT{Blocks \texttt{EPE-RISK-DOWNGRADE} if $T.\texttt{risk} < \rho_{\text{floor}}$}
\IF{$r_{\text{eff}} \in \{\texttt{medium}, \texttt{high}\}$}
  \STATE Verify \texttt{events.jsonl} hash chain; load Ed25519-signed approvals committed in audit log and apply signed revocations
  \STATE Require $\forall e \in T.\texttt{targets}$: matching revision $H_{\text{rev}}(e)$, risk coverage, independent reviewer ($r_{\text{eff}}=\texttt{medium}$), or $\texttt{kind}=\texttt{human}$ ($r_{\text{eff}}=\texttt{high}$)
\ENDIF
\STATE $\texttt{mapping\_ok} \leftarrow \text{ValidateASTMapping}(\mathcal{E}, \text{ParsePythonTree}(\mathcal{W}))$
\STATE $\texttt{arch\_ok} \leftarrow \text{ValidateArchitecture}(G, \mathcal{P}_{\text{arch}}, \text{ActiveApprovedADRs}(\mathcal{E})) \land \text{ValidateImportBoundaries}(\mathcal{W})$
\STATE $\texttt{scope\_ok} \leftarrow \text{CompareBaseline}(\mathcal{W}, T.\texttt{changeset}) \subseteq \texttt{write\_scope} \setminus \text{EffectiveProtected}(\mathcal{C}, T.\texttt{changeset})$
\FORALL{required check $k \in \mathcal{P}_{\text{verify}}.\texttt{checks}$ (fail-closed if empty)}
  \STATE Select latest UTC-timestamped evidence $E^*$ matching $(k.\texttt{id}, k.\texttt{command}, k.\texttt{version}, T.\texttt{id}, T.\texttt{changeset})$
  \STATE Recompute $v(E^*) \in \{\texttt{valid}, \texttt{stale}, \texttt{invalid}, \texttt{unknown}\}$ and ATMS $v_{\text{def}}$ over artifacts, policy hash, JUnit XML ($\ge 1$ passed, $0$ failed), runtime `(python, os)`, $\texttt{max\_age\_seconds}$, $\mathcal{M}_{\text{ver}}(\mathcal{W})$, and \texttt{evidence\_committed} audit entry
  \IF{$E^*$ missing \textbf{or} $E^*.\texttt{result} \neq \texttt{passed}$ \textbf{or} $v(E^*) \neq \texttt{valid}$}
    \STATE Append blocking diagnostic to $\mathcal{D}$
  \ENDIF
\ENDFOR
\RETURN $(\texttt{passed} \text{ if } \mathcal{D} = \emptyset \text{ else } \texttt{blocked}/\texttt{error}),\, \mathcal{D}$
\end{algorithmic}
\end{algorithm}

\section{Governance Ablation Detail}
\label{app:ablation}

Table~\ref{tab:rq2_ablation} reports the full conditional regime metrics across all 14 fault families ($N = 560$) plus the original 11-family subset ($N = 440$) and G1--G5 leave-one-out gate attribution; all rates carry two-sided 95\% Wilson score intervals.
\begin{table}[t]
  \centering
  \caption{Conditional governance safety and liveness over 560 randomized adversarial trials (14 families) and 60 valid workflows (seed 20260520), plus G1--G5 leave-one-out gate attribution.}
  \label{tab:rq2_ablation}
  \resizebox{\textwidth}{!}{%
  \begin{tabular}{lcccc}
    \toprule
    Governance Regime & FAR\textsubscript{14fam} (\%) $\downarrow$ [95\% CI] & FAR\textsubscript{11fam} (\%) $\downarrow$ & FRR (\%) $\downarrow$ [95\% CI] & Accuracy\textsubscript{14fam} (\%) $\uparrow$ \\
    \midrule
    B0: Self-Reported (Prompt) & 100.00 [99.32, 100.00] & 100.00 & 0.00 [0.00, 6.02] & 9.68 \\
    B1: Static AST \& Schema & 83.75 [80.47, 86.57] & 79.32 & 0.00 [0.00, 6.02] & 24.35 \\
    B2: Boolean Exec Gate & 98.39 [96.97, 99.15] & 97.95 & 0.00 [0.00, 6.02] & 11.13 \\
    B\textsubscript{git}: Minimal Git+Tag+Scope & 86.25 [83.15, 88.86] & 82.50 & 0.00 [0.00, 6.02] & 22.10 \\
    B3: Strong Git+CI+Static+CODEOWNERS & 49.46 [45.34, 53.59] & 57.50 & 0.00 [0.00, 6.02] & 55.32 \\
    \midrule
    EPE\textsubscript{unanch}: Unanchored Declared Risk & 32.50 [28.75, 36.49] & 14.09 & 0.00 [0.00, 6.02] & 70.65 \\
    EPE\textsubscript{floor}: Local Policy Risk Floor & 25.36 [21.93, 29.12] & 14.09 & 0.00 [0.00, 6.02] & 77.10 \\
    \textbf{EPE\textsubscript{full}: CI-Anchored 4D Epistemic} & \textbf{11.07 [8.73, 13.94]} & \textbf{14.09} & \textbf{0.00 [0.00, 6.02]} & \textbf{90.00} \\
    \midrule
    \multicolumn{5}{l}{\textit{Anchored EPE\textsubscript{full} decision decomposition ($N = 560$) and G1--G5 leave-one-out ($\Delta$FAR)}} \\
    \multicolumn{5}{l}{Definitive blocks: 88.93\% [86.06, 91.27] (498/560) \quad Conservative blocks: 0.00\% (0/560) \quad Silent accepts: 11.07\% (62/560)} \\
    \multicolumn{5}{l}{Leave-one-out FAR impact: G1 $+23.93$\,pp (134 uniq.) \;\; G5 $+22.32$\,pp (125) \;\; G3 $+21.07$\,pp (118) \;\; G4 $+9.11$\,pp (51) \;\; G2 $+7.14$\,pp (40)} \\
    \bottomrule
  \end{tabular}%
  }
\end{table}

\section{Governance Regime Baseline Definitions}
\label{app:baselines}

To isolate the causal contribution of each BTAE layer and deployment trust boundary, RQ2 compares eight governance configurations on identical repository states: (i) \textbf{B0 (Self-Reported / Prompt-Only)}, accepting frontmatter declarations and unverified completion claims; (ii) \textbf{B1 (Static AST \& Schema Only)}, enforcing JSON/YAML schemas, AST symbol mapping, and static forbidden-import rules without cryptographic approvals, ADR lifecycles, or evidence manifests; (iii) \textbf{B2 (Boolean Execution Gate)}, checking subprocess test exit codes without freshness or scope checks; (iv) \textbf{B\textsubscript{git} (Minimal Git Tree-Hash + Signed Tag)}, verifying a signed Ed25519 tag over \texttt{git write-tree}, passing exit status, and diff path \texttt{write\_scope}; (v) \textbf{B3 (Strong Practical Stack)}, uniting \texttt{B1} $\cup$ \texttt{B2} $\cup$ \texttt{B\textsubscript{git}} with CODEOWNERS protected-path enforcement and CI branch protection against local config/engine tampering; (vi) \textbf{EPE\textsubscript{unanch}}, local EPE trusting declared \texttt{T.risk} without risk floors; (vii) \textbf{EPE\textsubscript{floor}}, local EPE enforcing policy-derived blast-radius risk floors and global \texttt{protected\_paths}; and (viii) \textbf{EPE\textsubscript{full}}, full BTAE/EPE under anchored CI/read-only policy boundaries.

\section{Stratified Context Assembly Strata}
\label{app:strata}

Given a task $T$, fresh snapshot $G$, and item budget $B$, \texttt{resolve\_context} assembles a provenance-tagged bundle across the following four epistemic strata, summarized in Section~\ref{sec:method_context}:
\begin{enumerate}[leftmargin=1.8em,itemsep=1pt,topsep=3pt]
  \item \textbf{Stratum \texttt{C0} (Mandatory Charter \& Rules):} Always loads \texttt{AGENTS.md}, \texttt{docs/index.md}, and configured required policy sources first so global invariants cannot be trimmed silently.
  \item \textbf{Stratum \texttt{C1} (Target \& Graph-Adjacent Designs):} Adds the design document of every $e \in T.\texttt{target\_entities}$ plus active 1-hop EKG neighbors along incoming and outgoing edges, recording anchor entity, relation type, origin, and verification status.
  \item \textbf{Stratum \texttt{C2} (Mapped Code, Tests \& Lexical Matches):} Adds \texttt{implementation[]} source files and \texttt{verification[]} test files mapped to $T$'s targets, followed by active entities matching task-objective keywords (\texttt{confidence} $= 0.5$).
  \item \textbf{Stratum \texttt{C3} (Historical Decisions \& Task Evidence):} Appends Architecture Decision Records (\texttt{docs/architecture/decisions/}), archived designs (\texttt{freshness:\ historical}), and task-linked \texttt{evidence.json} summaries labeled with their current validity, strictly after \texttt{C0}--\texttt{C2}.
\end{enumerate}

\end{document}